\documentclass[aps,onecolumn,preprintnumbers,nofootinbib,superscriptaddress]{revtex4}
\usepackage{eurosym}
\usepackage{amsmath}
\usepackage{graphicx}
\usepackage{amsfonts}
\usepackage{array}
\usepackage{amsthm}
\usepackage{bm}
\usepackage{palatino}
\usepackage{mathpazo}
\usepackage{supertabular}
\usepackage{subfig}
\usepackage[breaklinks]{hyperref}
\usepackage[font=small,labelfont=bf,
justification=justified,
format=plain]{caption}
\usepackage[font=small,labelfont=bf,justification=justified]{caption}
\usepackage{color}
\usepackage{booktabs}
\usepackage{multirow}
\usepackage[labelfont=bf,labelsep=quad,justification=justified]{caption}
\newcommand{\be}{\begin{equation}}
\newcommand{\ee}{\end{equation}}
\newcommand{\ba}{\begin{eqnarray}}
\newcommand{\ea}{\end{eqnarray}}
\newcommand{\bal}{\begin{align}}
\newcommand{\eal}{\end{align}}

\newcommand{\bw}{\begin{widetext}}
\newcommand{\ew}{\end{widetext}}

\begin{document}

\title{Black hole surrounded by a dark matter halo in the M87 galactic center and its identification with shadow images}
%\title{ Black hole surrounded by Dark Matter Halo in the M87 Galactic Center and its Identification with Shadow Images  }
\author{Kimet Jusufi}
\email{kimet.jusufi@unite.edu.mk}
\affiliation{Physics Department, State University of Tetovo, Ilinden Street nn, 1200,
Tetovo, North Macedonia}
\affiliation{Institute of Physics, Faculty of Natural Sciences and Mathematics, Ss. Cyril
and Methodius University, Arhimedova 3, 1000 Skopje, North Macedonia}

\author{Mubasher Jamil}
\email{mjamil@zjut.edu.cn}
\affiliation{Institute for Theoretical Physics and Cosmology
Zhejiang University of Technology
Hangzhou, 310023 China}
\email{sumarna.haroon@sns.nust.edu.pk}\affiliation{Department of Mathematics, School of Natural
	Sciences (SNS), National University of Sciences and Technology
	(NUST), H-12, Islamabad, Pakistan}

\author{Paolo Salucci}
\email{salucci@sissa.it}
\affiliation{SISSA/ISAS, International School for Advanced Studies, Via Bonomea 265, 34136, Trieste, Italy}

\author{Tao Zhu}
\email{zhut05@zjut.edu.cn}
\affiliation{Institute for Theoretical Physics and Cosmology
Zhejiang University of Technology
Hangzhou, 310023 China}

\author{Sumarna Haroon}\email{sumarna.haroon@sns.nust.edu.pk}\affiliation{Department of Mathematics, School of Natural
	Sciences (SNS), National University of Sciences and Technology
	(NUST), H-12, Islamabad, Pakistan}
	
\begin{abstract}
In this paper we present a new black hole solution surrounded by dark 
matter halo in the galactic center using the  mass model of M87 and that 
coming from the  Universal Rotation Curve (URC) dark matter 
profile representing  family of spiral galaxies. In both cases the DM 
halo density is cored with a size $r_0$ and a central density $\rho_0$: 
$\rho(r)= \rho_0/(1+r/r_0)(1+(r/r_0)^2)$. Since $r_0\rho_0=120 $ M\textsubscript{\(\odot\)}/pc$^2$ \cite{Donato}, then by varying the central density one can 
reproduce the DM profile in any spiral. Using the Newman-Jains
 method we extend our solution to obtain a rotating black hole 
surrounded by dark matter halo.
We find that, the apparent shape of the shadow beside the black hole 
spin $a$, it also depends on the central density of the surrounded dark matter $\rho_0$. As a specific example we consider the galaxy M87, with a central density
$\rho_0=6.9\,\times 10^{6}$ M\textsubscript{\(\odot\)}/kpc$^3$ and a core radius $r_0=91.2$ kpc. In the case of M87, our analyses  show that the effect of dark matter on the size of the black hole shadow is almost negligible compared to the 
shadow size of the Kerr vacuum solution hence the angular diameter $ 42$ $\mu$as remains almost unaltered when the dark matter is considered.  For a small totally dark 
matter dominated spiral such as UGC 7232, we find similar effect of dark matter on the shadow images compared to the M87. However, in specific conditions having a core radius comparable to the black hole mass and dark matter with very high density, we show that the shadow images decreases compared to the Kerr vacuum black hole.  The effect of dark matter on the apparent shadow shape can shed some light in future 
observations as an indirect way to detect dark matter using the shadow 
images.

\end{abstract}
\maketitle
\tableofcontents

\section{Introduction}

Black holes (BHs) are the one of the most fascinating astrophysical objects which perform manifestations of extremely strong gravity such as formation of gigantic jets of particles and disruption of neighboring stars. From theoretical perspective, BHs serve as a lab to test various predictions of theories of quantum gravity such as the Hawking radiation. Recently, the Event Horizon Telescope (EHT) Collaboration announced their first results concerning the detection of an event horizon of a supermassive black hole at the center of a neighboring elliptical M87 galaxy \cite{m87}. The bright accretion disk surrounding the black hole appears distorted due to the phenomenon of gravitational lensing. The region of accretion disk behind the black hole also gets visible due to bending of light by black hole. The shadow image helps in understanding the geometrical structure of the event horizon and the speed of rotation of the black hole.

In this paper, we consider a scenario of a BH surrounded by a halo containing mostly dark matter. We ignore the effects of baryonic matter and may consider it as a separate study.  For dark matter, we assume a Burkert halo (or the Universal Rotation Curve) profile of normal spirals which point to the core to have fixed radius and density \cite{saluci1,Burkert,saluci2,Donato,lapi,saluci3,saluci4}. We like to understand the effects of dark matter over the shadows of black hole immersed in the halo in a family of spiral galaxies. As a particular example we shall consider the M87 black hole located at the galactic center and a small spiral totally dominated by dark matter. Thus it will be interesting to see if current or future astronomical observations can detect the presence of dark matter using the black hole shadow.

Theoretically, the shadow cast by the black hole horizon is studied as null geodesics and the existence of a photon sphere. The incoming photons are trapped in an unstable circular orbit ($r=3M$ for a Schwarzschild BH). Occasionally the photons are perturbed and diverted towards the observer. The shadow of a BH is characterized by the celestial coordinates $\alpha$ and $\beta$ which are plotted for different values of the BH parameters such as mass and spin. In literature, the theory of black hole shadows is well-developed and is under investigation for decades. After the seminal work of Synge \cite{Synge} and \cite{Luminet} on the apparent shape of a spherically symmetric black hole  and the appearance Schwarzschild black hole, some notable results of BH shadows are already discussed in the literature, to list a few: The shadow of a Kerr black hole was studied by Bardeen \cite{Bardeen}, shadow of Kerr-Newman black holes \cite{devaries}, naked singularities with deformation parameters \cite{hioki},  Kerr-Nut spacetimes \cite{abu}, while shadows of black holes in Chern-Simons modified gravity, Randall-Sundrum braneworlds, and Kaluza-Klein rotating black holes have been studied in \cite{Amarilla1,Amarilla2,Amarilla3}, shadow of Kerr-Perfect fluid dark matter BH \cite{pfdm,Hou:2018bar,Xu:2018mkl,Konoplya:2019sns} and rotating global monopoles  with dark matter \cite{gm} and many other interesting studies concerning the Kerr-like wormholes as well as traversable wormholes and many others interesting studies \cite{shaikh1,shaikh2,G1,G2,A2,Amir1}. Some authors have also tried to test theories of gravity by using the observations obtained from shadow of Sgr A* \cite{Bambi,Bro,B1,M}; Einstein-dilaton-Gauss-Bonnet BH \cite{gb}; Konoplya-Zhidenko BH \cite{kz}; Einstein-Maxwell-Chern-Simon BH \cite{cs}; Kerr-Newman-Kasuya BH \cite{knk};  Kerr de Sitter BH \cite{kds}; Kerr-MOG BH \cite{mog}; rotating regular BH \cite{rr}; Kerr-Sen BH \cite{ks}; non-commutative BHs \cite{nc}; naked singularities \cite{ns,Gyulchev:2019tvk}, etc. 

The plan of the paper is as follows: In Sec.II, we use the URC dark matter profile to obtain the radial function of a spherically symmetric spacetime. In Sec. III, we find a spherically symmetric black hole metric surrounded by dark matter  halo. In Sec. IV, we Newman-Jains method to find a rotating black hole with dark matter effects. In Secs. V and VI we study null geodesics and circular orbits, respectively. In Secs. VII and VIII we investigate the shadow images and  radius distortion. Finally in Sec. IX we comment on our results. We shall use the natural units $G=c=\hbar=1$ through the paper.

\section{The universal rotation curve and the spacetime metric of the dark matter halo}
%From numerical simulations in the region closer to the center where the Universal Rotation Curve (URC) dark matter profile.
It is well known that the  actual density  halo distribution around galaxies is well represented  by the Burkert profile  as first  proposed  by Salucci and Burkert \cite{saluci1,Burkert}. This distribution has been confirmed by the deep  investigation of   the family of spiral {\it coadded } rotation curves derived from 1000 individual RCs (see, Salucci et al \cite{saluci2}) and of a large number of individual RC's (see Donato et al \cite{Donato}). An even bigger number of RCs  has also confirmed this density distribution (see Lapi et al \cite{lapi}), while for a review see \cite{saluci3}.  This density profile is also known as the URC dark matter profile and can be given by the following relation \cite{saluci1}
\begin{eqnarray}
\rho(r)=\frac{\rho_0 r^3_0}{(r+r_0)(r^2+r_0^2)},
\end{eqnarray}
with the mass profile of the dark matter
galactic halo given by
\begin{equation}
M_{DM}\left( r\right) =4\pi \int_{0}^{r}\rho
\left( r'\right) r^{'2}dr'.
\end{equation}

Solving the last integral we obtain
\begin{align}
      M_{DM}\left( r\right) =\pi \rho_0 r_0^3\left[ \ln(1+\frac{r^2}{r_0^2})+2 \ln(1+\frac{r}{r_0})-2 \arctan(\frac{r}{r_0}) \right]
\end{align}

From the last equation one can find the tangential velocity $%
v_{tg}^{2}\left( r\right) =M_{DM}(r)/r$ of a test particle moving
in the dark halo in spherical symmetric space-time as follows
\begin{align}
      v_{tg}\left( r\right) =\sqrt{\frac{\pi \rho_0 r_0^3}{r}}\left[ \ln(1+\frac{r^2}{r_0^2})+2 \ln(1+\frac{r}{r_0})-2 \arctan(\frac{r}{r_0}) \right]^{1/2}.
\end{align}

In this section, we derive the space-time geometry for pure dark matter. To do so, let us consider a static and spherically symmetric spacetime ansatz with pure dark matter in Schwarzschild coordinates can be written as follows 
\begin{equation}
\mathrm{d}s^{2}=-f(r)\mathrm{d}t^{2}+\frac{\mathrm{d}r^{2}}{g(r)}+r^{2}\left(
\mathrm{d}\theta ^{2}+\sin ^{2}\theta \mathrm{d}\phi ^{2}\right),  \label{5}
\end{equation}
in which $f(r)$ and $g(r)$ are known as the redshift and shape
functions, respectively. Given the tangential velocity, one can calculate the radial function $f(r)$ by the following equation \cite{Xu:2018wow}
\begin{align}
      v_{tg}^{2}\left( r\right) =\frac{r}{\sqrt{f(r)}}\frac{d \sqrt{f(r)}}{dr}=r \frac{d\ln(\sqrt{f(r)})}{dr}.
\end{align}

Assuming a spherically symmetric solution, i.e. $f(r)=g(r)$, and solving the last equation we find
\begin{equation}
f(r)=\left(1+\frac{r^2}{r_0^2}\right)^{-\frac{2 \rho_0 r_0^3 \pi}{r}(1-\frac{r}{r_0})}\left(1+\frac{r}{r_0} \right)^{-\frac{4 \rho_0 r_0^3 \pi}{r}(1+\frac{r}{r_0})}\exp\left(\frac{4 \rho_0 r_0^3 \pi \arctan(\frac{r}{r_0})(1+\frac{r}{r_0})}{r}   \right)
\end{equation}

In the specific case, we shall consider the galaxy M87, with the central density given by  $\rho_0=6.9\,\times 10^{6} $M\textsubscript{\(\odot\)}/kpc$^3$ with a core radius of the galaxy M87 given by $r_0=91.2$ kpc. It's worth noting that the central densities depends upon the core radius $r_0$. Using the mass of the black hole M87  estimated as $M_{BH}=6.5 \,\times 10^{9} $M\textsubscript{\(\odot\)} we can express the central density as $\rho_0=0.001$ in units of $M_{BH}/$kpc$^3$.  At this stage it is convenient to introduce a new constant $k$ defined by the following relation $k=\rho_0 r_0^3$. In particular for M87 we find $k=805$ in units of black hole M87 mass. 

If the dark matter is absent, i.e. $k =0$, our solution for the radial function $f(r)$ reduces to the following expected relation
\begin{equation}
\lim_{k \to 0}f(r)=\lim_{k  \to 0}\left\lbrace\left(1+\frac{r^2}{r_0^2}\right)^{-\frac{2 \,k \pi}{r}(1-\frac{r}{r_0})}\left(1+\frac{r}{r_0} \right)^{-\frac{4\, k  \pi}{r}(1+\frac{r}{r_0})}\exp\left(\frac{4\,k \pi \arctan(\frac{r}{r_0})(1+\frac{r}{r_0})}{r}   \right) \right\rbrace=1
\end{equation}

\section{ Black holes in dark matter halo}
We now shall generalize our solution to consider black holes surrounded by dark matter halo. To do so, first let us review breafly the result obtained in Ref. \cite{Xu:2018wow}. From the pure dark matter space-time it was argued that one can obtain the space-time metric of a black hole surrounded by dark matter halo. As a special case one can recover the Schwarzschild metric when dark matter is not absent. From the Einstein field equation we have
\begin{equation}
R^{\nu}_{~\mu}-\dfrac{1}{2}\delta^{\nu}_{~\mu}R=\kappa^{2}{T^{\nu}_{~\mu}}^{DM}.
\label{SPBH7}
\end{equation}

Here ${T^{\nu}_{~\mu}}^{DM}=diag[-\rho,p_{r},p,p]$ are the non-zero energy-momentum corresponding to the pure dark matter space-time metric, given by \cite{Xu:2018wow}
\begin{equation}
\kappa^{2}{T^{t}_{t}}^{DM}=g(r)(\dfrac{1}{r}\dfrac{g^{'}(r)}{g(r)}+\dfrac{1}{r^{2}})-\dfrac{1}{r^{2}},$$$$
\kappa^{2}{T^{r}_{r}}^{DM}=g(r)(\dfrac{1}{r^{2}}+\dfrac{1}{r}\dfrac{f^{'}(r)}{f(r)})-\dfrac{1}{r^{2}},$$$$
\kappa^{2}{T^{\theta}_{\theta}}^{DM}=\kappa^{2}{T^{\phi}_{\phi}}^{DM}=\dfrac{1}{2}g(r)(\dfrac{f^{''}(r)f(r)-f^{'2}(r)}{f^{2}(r)}+\dfrac{1}{2}\dfrac{f^{'2}(r)}{f^{2}(r)}+\dfrac{1}{r}(\dfrac{f^{'}(r)}{f(r)}+\dfrac{g^{'}(r)}{g(r)})+\dfrac{f^{'}(r)g^{'}(r)}{2f(r)g(r)}).
\end{equation}

One way to include the black hole in our metric is by treating the dark matter as part of the general energy-momentum tensor $\mathcal{T}_{\mu \nu}=T_{\mu \nu}+{T_{\mu \nu}}^{DM}$.  At this point, we emphasize that for the Schwarzschild black hole we need to take into account only the energy-momentum tensor related to the dark matter since the energy-momentum tensor of the Schwarzschild black hole is zero, $T_{\mu\nu}=0$. This we guess our space-time metric as follows
\begin{equation}
ds^{2}=-(f(r)+F_{1}(r))\,dt^{2}+\frac{dr^{2}}{g(r)+F_{2}(r)}+r^{2}(d\theta^{2}+\sin^{2}\theta d\phi^{2}).
\end{equation}
For convenience let us rewrite these coefficient functions as follows
\begin{equation}
F(r)=f(r)+F_{1}(r),$$$$
G(r)=g(r)+F_{2}(r).
\end{equation}

The Einstein field equation can now be written as
\begin{equation}
R^{\nu}_{~\mu}-\dfrac{1}{2}\delta^{\nu}_{~\mu}R=\kappa^2(T^{\nu}_{~\mu}+{T^{\nu}_{~\mu}}^{DM}).
\end{equation}
With the help of our space-time metric and Einstein field equations yields
\begin{equation}
(g(r)+F_{2}(r))(\dfrac{1}{r^{2}}+\dfrac{1}{r}\dfrac{g^{'}(r)+F^{'}_{2}(r)}{g(r)+F_{2}(r)})=g(r)(\dfrac{1}{r^{2}}+\dfrac{1}{r}\dfrac{g^{'}(r)}{g(r)}),$$$$
(g(r)+F_{2}(r))(\dfrac{1}{r^{2}}+\dfrac{1}{r}\dfrac{f^{'}(r)+F^{'}_{1}(r)}{f(r)+F_{1}(r)})=g(r)(\dfrac{1}{r^{2}}+\dfrac{1}{r}\dfrac{f^{'}(r)}{f(r)}).
\end{equation}

In terms of these relations space-time metric including a black hole in dark matter halo gives \cite{Xu:2018wow}
\begin{equation}
ds^{2}=-\exp[\int \dfrac{g(r)}{g(r)-\dfrac{2GM}{r}}(\dfrac{1}{r}+\dfrac{f^{'}(r)}{f(r)})dr-\dfrac{1}{r} dr]dt^{2}+(g(r)-\dfrac{2GM}{r})^{-1}dr^{2}
+r^{2}(d\theta^{2}+\sin^{2}\theta d\phi^{2}).
\end{equation}

In the special case when dark matter is absent i.e., $f(r)=g(r)=1$, the indefinite integral results with a constant
\begin{equation}
F_{1}(r)+f(r)=\exp[\int \dfrac{g(r)}{g(r)+F_{2}(r)}(\dfrac{1}{r}+\dfrac{f^{'}(r)}{f(r)})dr-\dfrac{1}{r} dr]=1-\dfrac{2M}{r},
\end{equation}
with $M$ being the black hole mass. In other words, we end up with the Schwarzschild black hole space-time. Finally we can interpret the space-time Eq.(15) as a Schwarzschild black hole surrounded by dark matter halo. Given the dark matter density profile one can obtain the corresponding space-time. Based on our assumption that $f(r)=g(r)$, we obtain $F_{1}(r)=F_{2}(r)=-2M/r$, hence the black hole space-time is finally written as
\begin{eqnarray}
ds^{2}=-F(r)dt^{2}+\frac{dr^2}{G(r)}+H(r)(d\theta^2+\sin^2\theta d\phi^2)
\end{eqnarray}
where $H(r)=r^2$. The black hole space-time metric coefficient functions are given by
\begin{equation}
F(r)=G(r)=\left(1+\frac{r^2}{r_0^2}\right)^{-\frac{2 \,k \pi}{r}(1-\frac{r}{r_0})}\left(1+\frac{r}{r_0} \right)^{-\frac{4\, k  \pi}{r}(1+\frac{r}{r_0})}\exp\left(\frac{4\,k \pi \arctan(\frac{r}{r_0})(1+\frac{r}{r_0})}{r}   \right)- \dfrac{2M}{r}.
\end{equation}

\section{Rotating black holes in dark matter halo}
We shall now generalize our spherical symmetric black hole to rotational black hole surrounded by dark matter halo based on the Newman-Jains method.  Following the standard formalism, firstly we transform Boyer-Lindquist (BL) coordinates $(t,r,\theta,\phi)$ to Eddington-Finkelstein (EF) coordinates $(u,r,\theta,\phi)$. One can obtain such coordinates by introducing the following transformation
\begin{eqnarray}\label{eq18}
dt&=&du+\frac{dr}{\sqrt{F(r)G(r)}},
\end{eqnarray}

It is convenient to rewrite this metric in terms of null tetrads as
\begin{eqnarray}
g^{\mu{\nu}}=-l^{\mu}n^{\nu}-l^{\nu}n^{\mu}+m^{\mu}\overline{m}^{\nu}+m^{\nu}\overline{m}^{\mu},
\end{eqnarray}
and the null tetrads defined by
\begin{eqnarray}
l^{\mu}&=&\delta^{\mu}_{r},\\
n^{\mu}&=& \delta^{\mu}_{u}-\frac{1}{2}F(r)\delta^{\mu}_{r},\\
m^\mu&=&\frac{1}{\sqrt{2H}}\left(\delta^{\mu}_{\theta}+\frac{\dot{\iota}}{\sin\theta}\delta^{\mu}_{\phi}\right),\\
\overline{m}^\mu&=&\frac{1}{\sqrt{2 H}}\left(\delta^{\mu}_{\theta}-\frac{\dot{\iota}}{\sin\theta}\delta^{\mu}_{\phi}\right)
\end{eqnarray}

It's worth noting that the null tetrads are chosen such that $m^\mu$ and $\bar{m}^\mu$ are complex, thus for example $\bar{m}^\mu$ is complex conjugate of $m^\mu$. By construction these vectors satisfy the conditions for normalization, orthogonality and isotropy as
\begin{eqnarray}
l^{\mu}l_{\mu}=n^{\mu}n_{\mu}=m^{\mu}m_{\mu}=\bar{m}^{\mu}\bar{m}_{\mu}=0,\\
l^{\mu}m_{\mu}=l^{\mu}\bar{m}_{\mu}=n^{\mu}m_{\mu}=n^{\mu}\bar{m}_{\mu}=0,\\
-l^{\mu}n_{\mu}=m^{\mu}\bar{m}_{\mu}=1.
\end{eqnarray}
According to the Newman--Janis prescription we can write
\begin{equation}
{x'}^{\mu} = x^{\mu} + ia (\delta_r^{\mu} - \delta_u^{\mu})
\cos\theta \rightarrow \\ \left\{\begin{array}{ll}
u' = u - ia\cos\theta, \\
r' = r + ia\cos\theta, \\
\theta' = \theta, \\
\phi' = \phi. \end{array}\right.
\end{equation}
with $a$ being the spin parameter. Furthermore the null tetrad vectors $Z^a$ transform
according to the relation  $Z^\mu = ({\partial x^\mu}/{\partial {x^\prime}^\nu}) {Z^\prime}^\nu $, yielding
\begin{eqnarray}
l'^{\mu}&=&\delta^{\mu}_{r},\\
n'^{\mu}&=&\sqrt{\frac{B}{A}}\delta^{\mu}_{u}-\frac{1}{2}B\delta^{\mu}_{r},\\\label{e11}
m'^{\mu}&=&\frac{1}{\sqrt{2\,\Sigma}}\left[(\delta^{\mu}_{u}-\delta^{\mu}_{r})\dot{\iota}{a}\sin\theta+\delta^{\mu}_{\theta}+\frac{\dot{\iota}}{\sin\theta}\delta^{\mu}_{\phi}\right],\\
\overline{m}'^{\mu}&=&\frac{1}{\sqrt{2\,\Sigma}}\left[(\delta^{\mu}_{u}-\delta^{\mu}_{r})\dot{\iota}{a}\sin\theta+\delta^{\mu}_{\theta}+\frac{\dot{\iota}}{\sin\theta}\delta^{\mu}_{\phi}\right],
\end{eqnarray}
where we have assumed that the functions $(G(r), F(r), H(r))$ transform to $(A(a,r,\theta), B(a,r,\theta), \Sigma(a,r,\theta))$ (see \cite{Azreg-Ainou:2014pra}). Having defined the null tetrad vectors one can construct the contravariant components of our new metric in terms of the following relations
\begin{eqnarray}\notag
g^{uu}&=&\frac{a^{2}\sin^{2}\theta}{\Sigma},\   \ g^{u\phi}=\frac{a}{\Sigma}, \   \ g^{ur}=1-\frac{a^{2}\sin^2\theta}{\Sigma},\\\notag
g^{rr}&=&\mathcal{F}+\frac{a^{2}\sin^{2}\theta}{\Sigma},\   \ g^{r\phi}=-\frac{a}{\Sigma},\   \ g^{\theta\theta}=\frac{1}{\Sigma},\\
g^{\phi\phi}&=&\frac{1}{\Sigma\sin^2\theta}.
\end{eqnarray}

Note $\Sigma=r^2+a^2 \cos^2\theta$, and $\mathcal{F}$  is some function of $r$ and $\theta$. The  metric is found as follows
\begin{eqnarray}
ds^2=-\mathcal{F} du^2-2dudr+2a\sin^2\theta\left(\mathcal{F}-1\right)du{d\phi}+2a\sin^2drd\phi+\Sigma d\theta^2+\sin^2\theta\left[\Sigma+a^2\left(2-\mathcal{F}\right)\sin^2\theta\right]d\phi^2.
\end{eqnarray}
Furthermore we can rewrite our black hole solution in terms of old coordinates by introducing the following transformations
\begin{eqnarray}
du=dt- \frac{a^2+r^2}{\Delta}dr,\,\,\,\,\, d\phi=d\varphi- \frac{a}{\Delta}dr,
\end{eqnarray}
where $\Delta$ is defined by
\begin{equation}
\Delta=r^{2}G(r)+a^{2}=r^2\left(1+\frac{r^2}{r_0^2}\right)^{-\frac{2 \,k \pi}{r}(1-\frac{r}{r_0})}\left(1+\frac{r}{r_0} \right)^{-\frac{4\, k  \pi}{r}(1+\frac{r}{r_0})}\exp\left(\frac{4\,k \pi \arctan(\frac{r}{r_0})(1+\frac{r}{r_0})}{r}   \right)- 2Mr+a^2,
\label{KBH2}
\end{equation}
with $G(r)=F(r)=f(r)+F_{1}(r)=g(r)+F_{2}(r)$. Note that the first terms in the last equation encodes the dark matter halo, while the second and third term depends on the black hole mass and angular momentum parameter, respectively. 
The rotational black hole space-time metric surrounded by dark matter halo is
%\begin{eqnarray}\label{3}\notag
%ds^2 & = & -\left(\frac{\Delta-a^2 \sin^2 \theta}{\Sigma}\right)dt^2
%-2 a  \sin^2 \theta \left(1-\frac{\Delta-a^2 \sin^2 \theta}{\Sigma}   \right) dt d\varphi+\frac{\Sigma}{\Delta}dr^2+\Sigma d\theta^2\\
%&+& \sin^2 \theta \left[ \Sigma+a^2 \sin^2\theta \left(2-\frac{\Delta-a^2 \sin^2 \theta}{\Sigma}\right)\right]d\varphi^2\label{metric}
%\end{eqnarray}
\begin{equation}
ds^2=-\left(1-\frac{2\Upsilon(r) r}{\Sigma}\right)dt^2+\frac{\Sigma}{\Delta}dr^2+\Sigma d\theta^2-2 a \sin^2\theta \frac{2\Upsilon(r) r}{\Sigma}dt d\varphi+\sin^2\theta \left[\frac{(r^2+a^2)^2-a^2\Delta \sin^2\theta}{\Sigma} \right] d\varphi^2\label{metric}
\end{equation}
where we have introduced
 \begin{equation}
\Upsilon(r)=\frac{r (1-G(r))}{2}.
\end{equation}

The last equation represents the space-time metric of rotational black holes surrounded by URC dark matter halo. In the Appendix A we proof that our spherical symmetric solution generated from Newman-Janis algorithm does satisfy the Einstein field equations.

 \begin{figure}[h!]
 \includegraphics[width=0.47\textwidth]{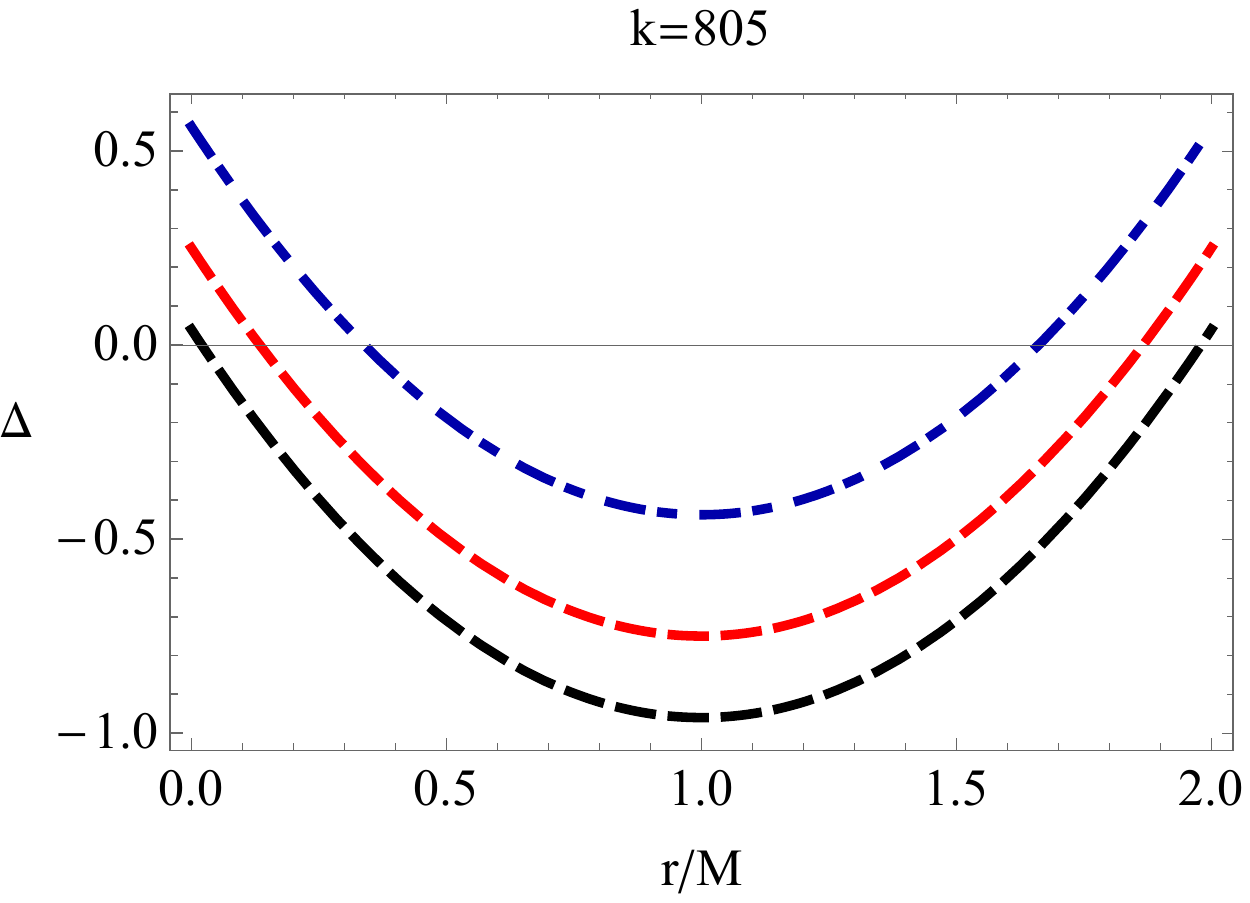}
\includegraphics[width=0.47\textwidth]{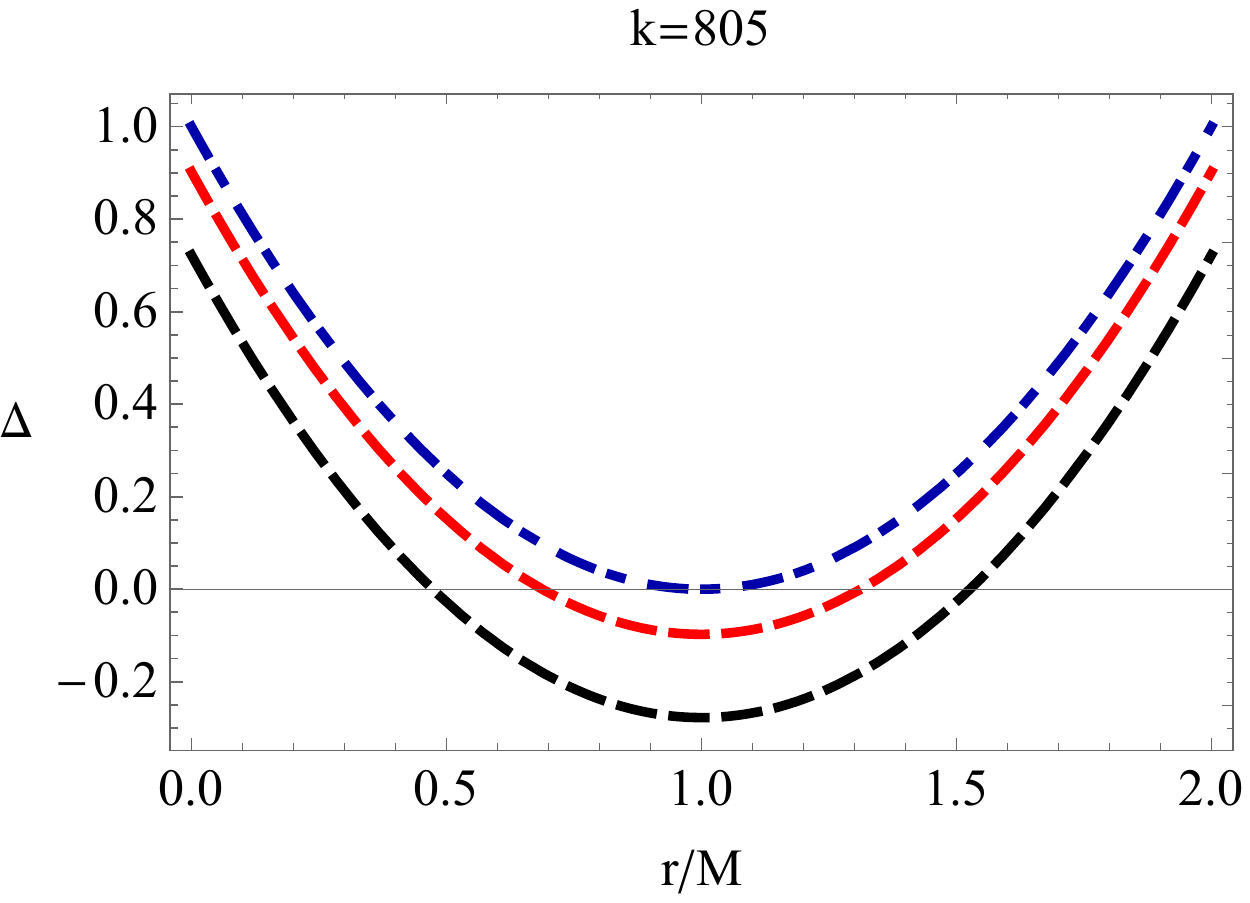}
  \caption{\label{figure2} Variation of $\Delta$ as a function of $r$, for a fixed value of $r_0=91.2$ kpc, $k=805$ in units of $M_{BH}/$kpc$^3$. Left panel: We use $a=0.2$ (black curve), $a=0.5$ (red curve),  $a=0.75$ (blue curve), respectively. Right panel: We use $a=0.85$ (black curve), $a=0.95$ (red curve),  $a=1$ (blue curve), respectively.}
  \end{figure}
  
   \begin{figure}[h!]
 \includegraphics[width=0.47\textwidth]{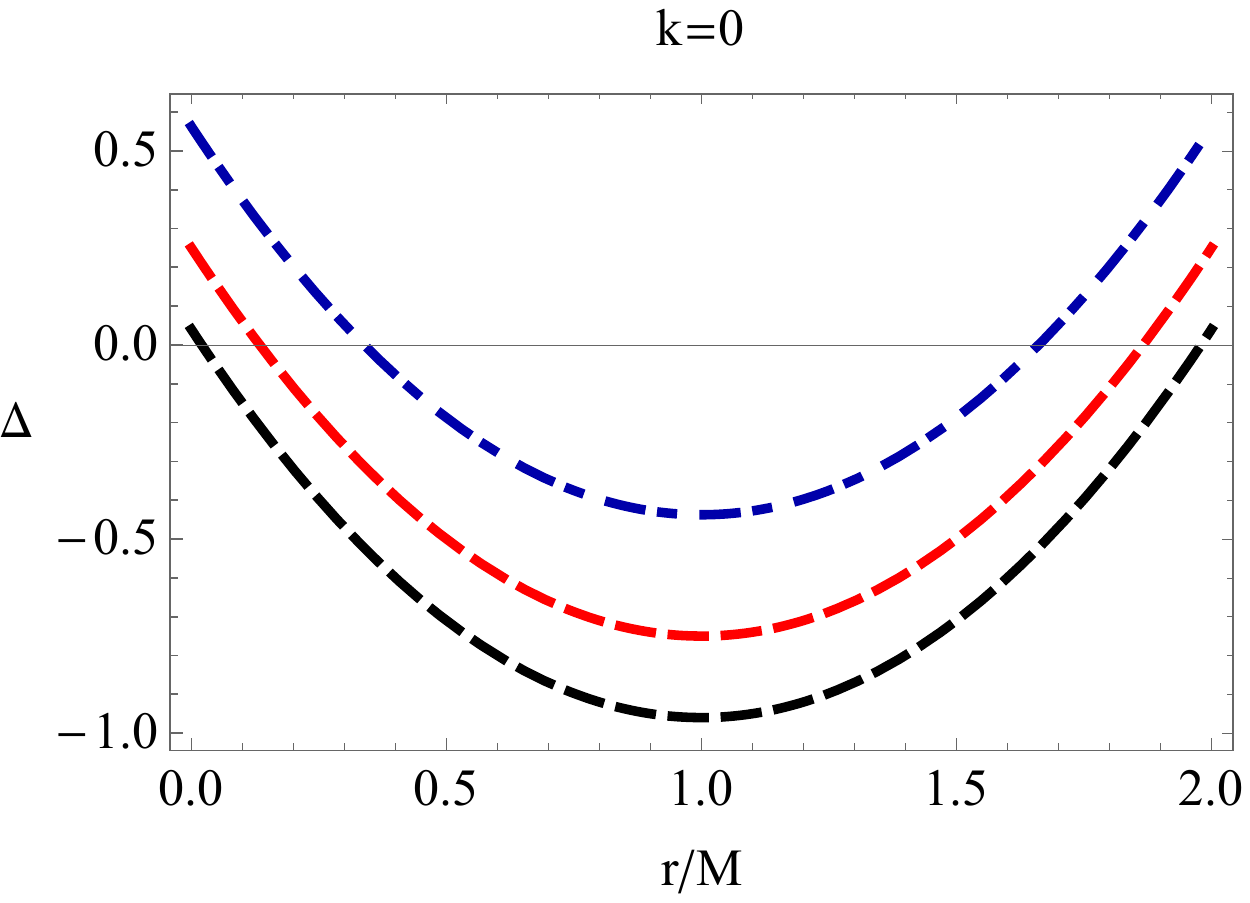}
\includegraphics[width=0.47\textwidth]{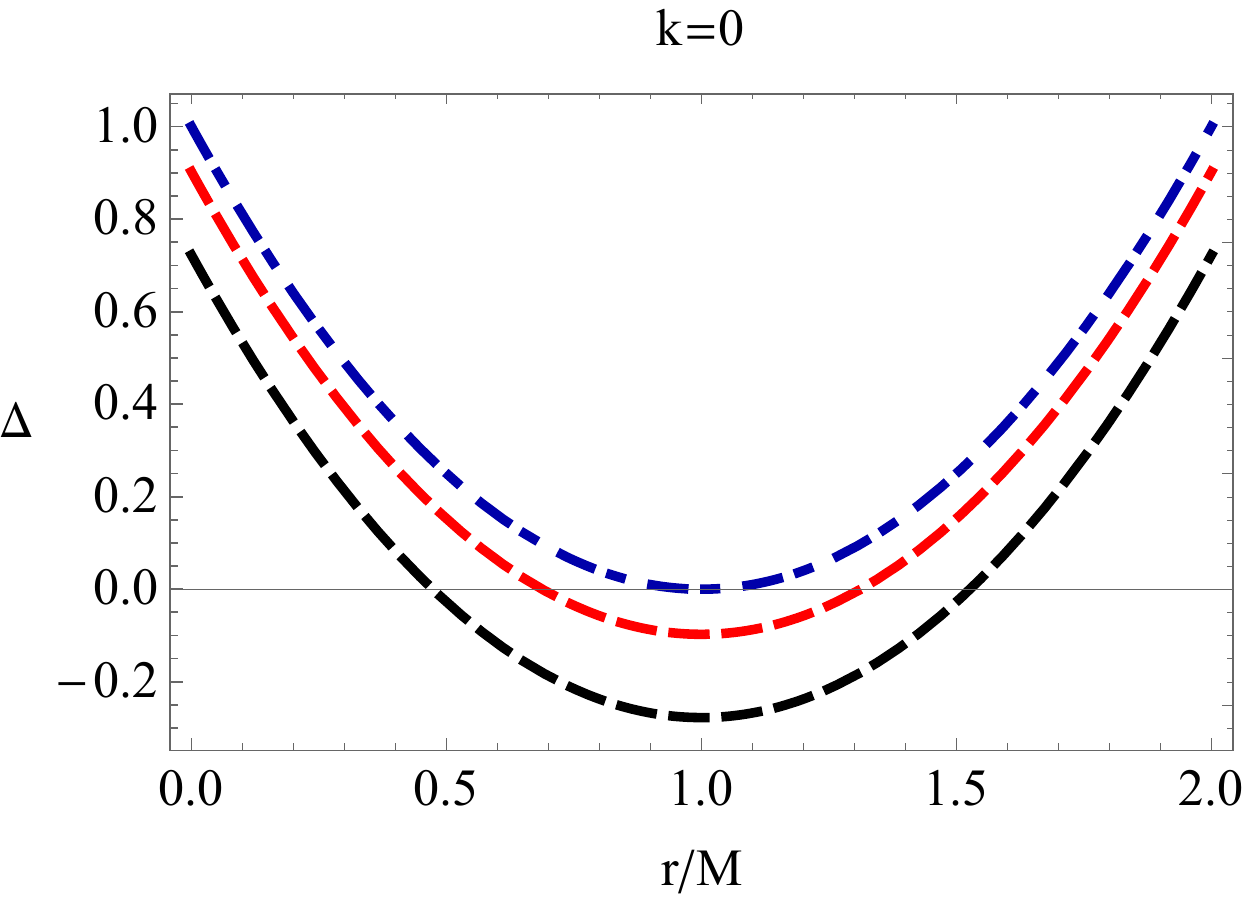}
  \caption{\label{figure2} Variation of $\Delta$ as a function of $r$, with no dark matter halo, i.e, $\rho_0=0$. Left panel: We use $a=0.2$ (black curve), $a=0.5$ (red curve),  $a=0.75$ (blue curve), respectively. Right panel: We use $a=0.85$ (black curve), $a=0.95$ (red curve),  $a=1$ (blue curve), respectively.}
  \end{figure}
  
  \begin{figure}[h!]
     \includegraphics[width=0.47\textwidth]{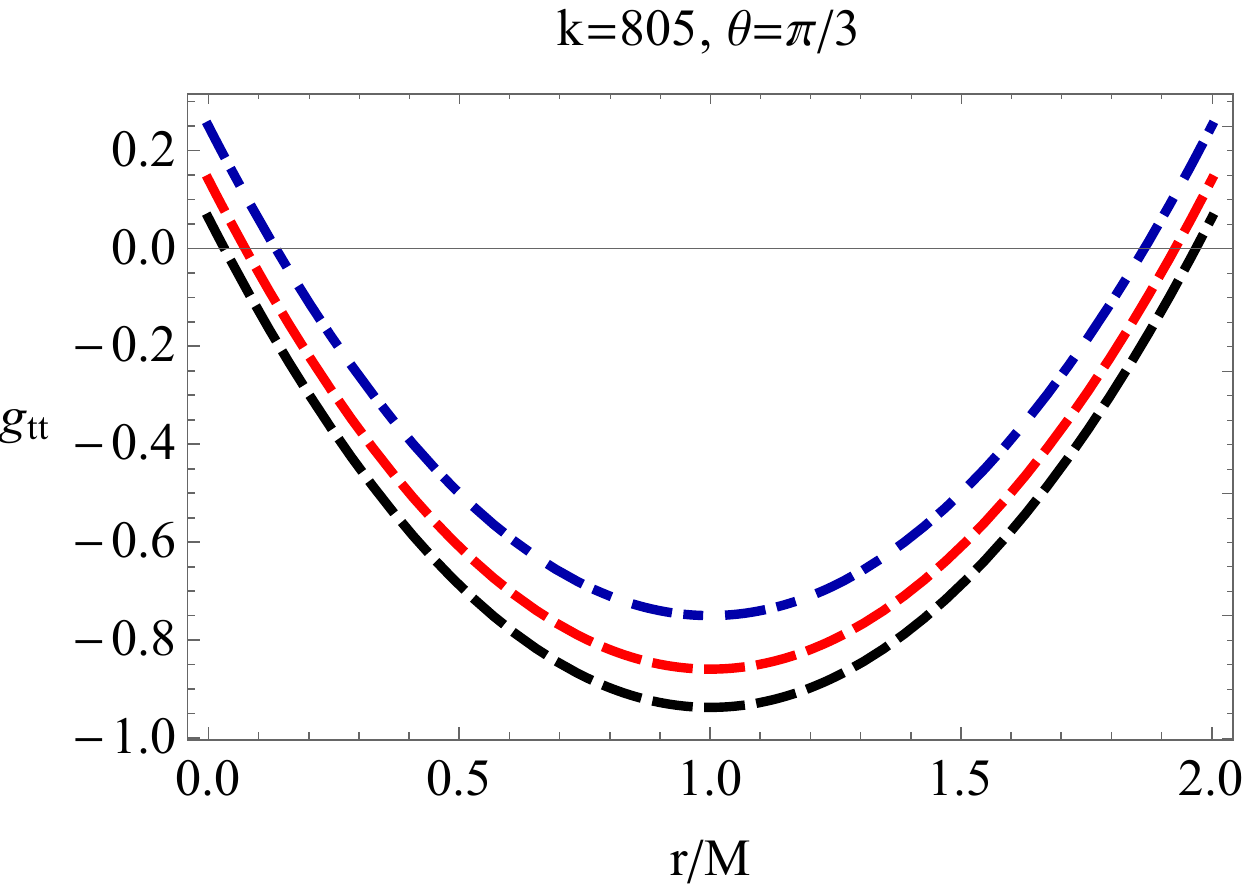}
          \includegraphics[width=0.47\textwidth]{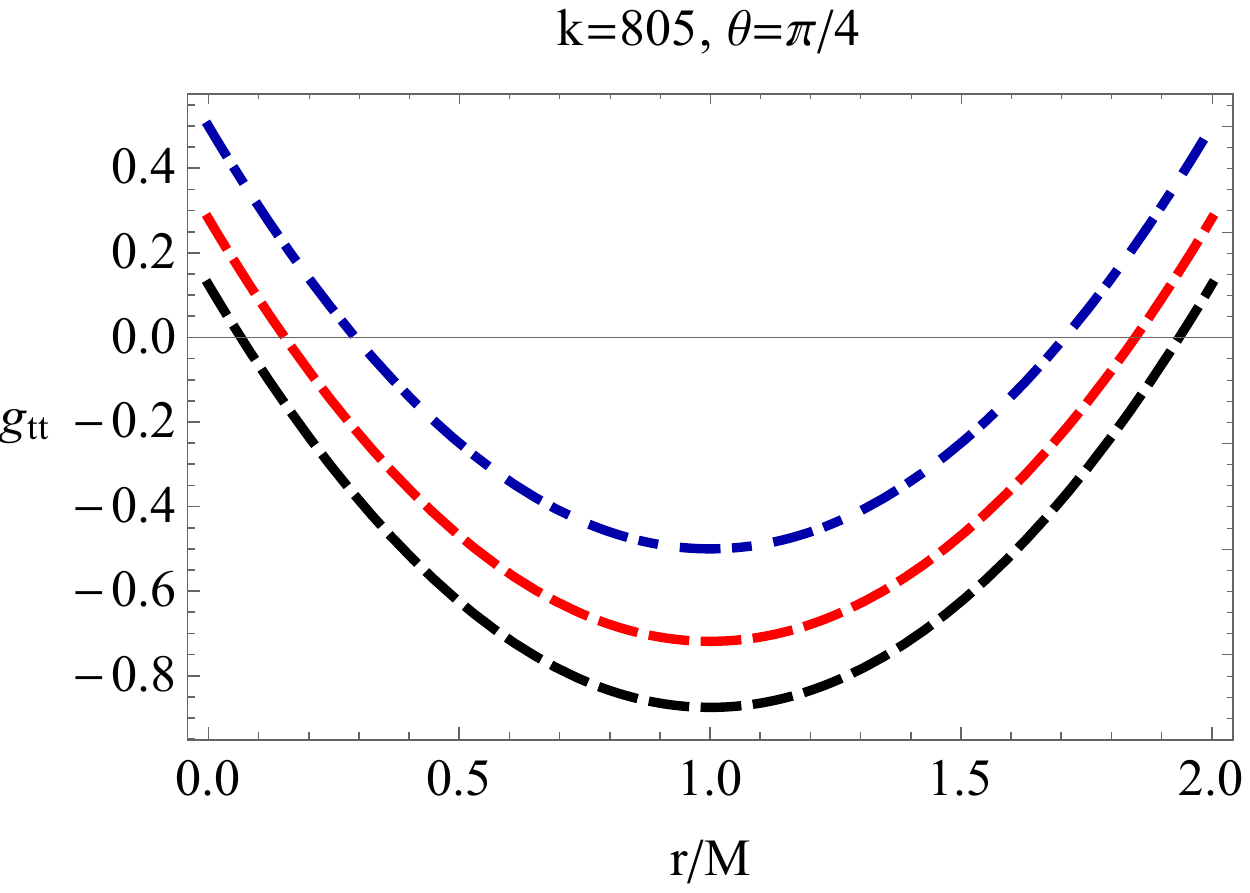}
  \caption{\label{figure2} Variation of $g_{tt}$ as a function of $r$, for a fixed value of $r_0=91.2$ kpc, and $k=805$ in units of $M_{BH}/$kpc$^3$.  We use $a=0.5$ (black curve), $a=0.75$ (red curve),  $a=1$ (blue curve), in both plots respectively. }
  \end{figure}
  
   \begin{figure}[h!]
     \includegraphics[width=0.47\textwidth]{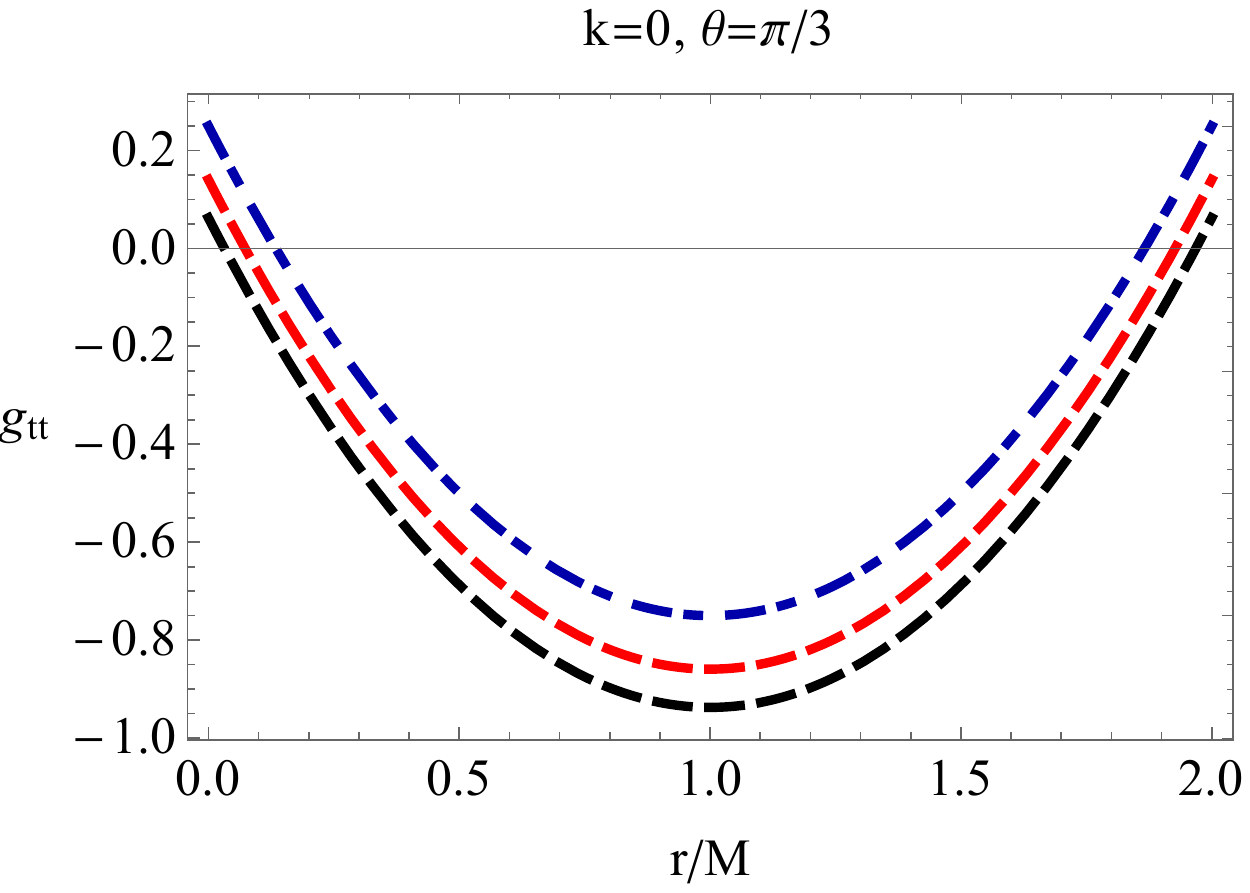}
          \includegraphics[width=0.47\textwidth]{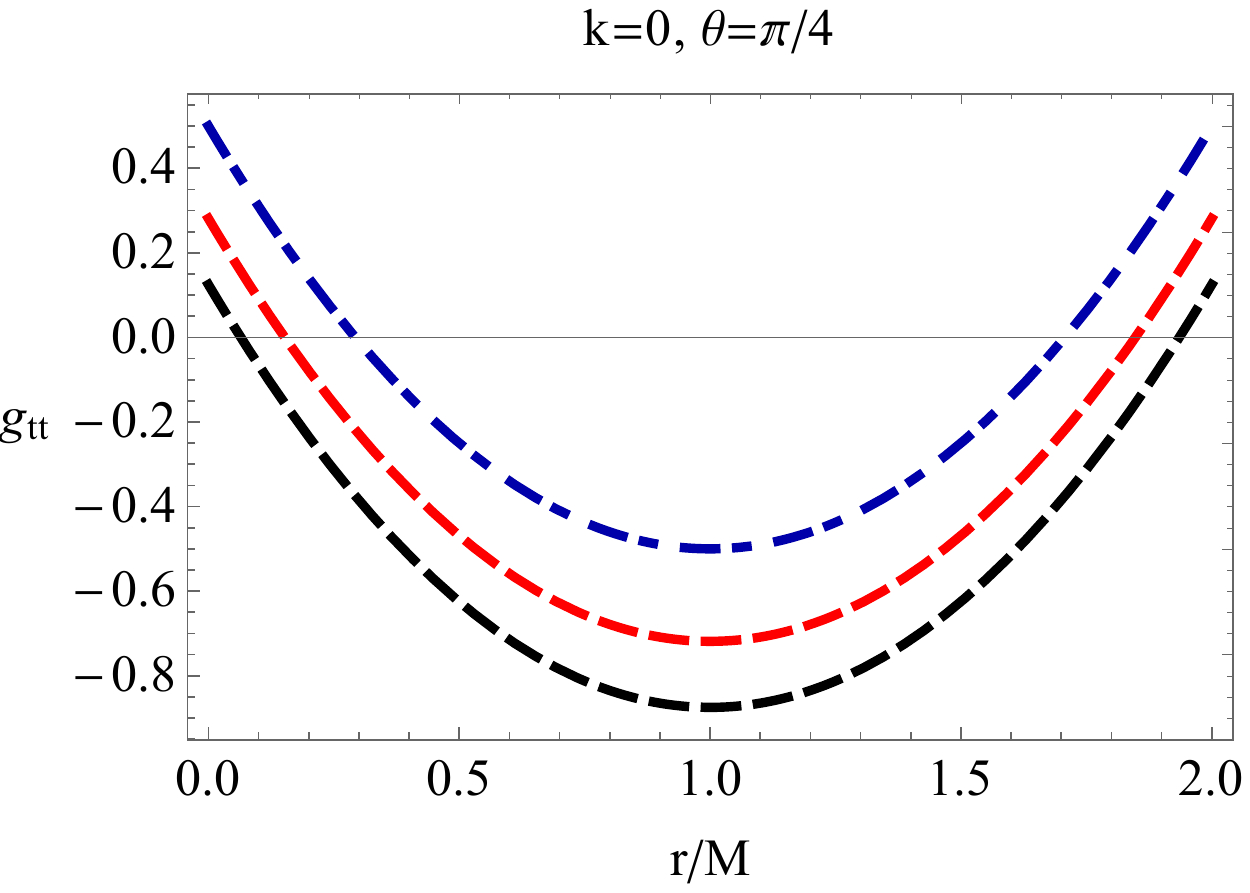}
  \caption{\label{figure2} Variation of $g_{tt}$ as a function of $r$,  with no dark matter halo, i.e, $\rho_0=0$. We use $a=0.5$ (black curve), $a=0.75$ (red curve),  $a=1$ (blue curve), in both plots respectively. }
  \end{figure}
  
\newpage
\subsection{Shape of Ergoregion}
Let us now proceed to study the shape of the ergoregion of our black hole metric given by \eqref{metric}. In particular we shall be interested to plot the shape of the ergoregion in the $xz$-plane. Recall that the horizons of the black hole can be found by solving $\Delta=0$, i.e., 
\begin{equation}
r^2\left(1+\frac{r^2}{r_0^2}\right)^{-\frac{2 \,k \pi}{r}(1-\frac{r}{r_0})}\left(1+\frac{r}{r_0} \right)^{-\frac{4\, k  \pi}{r}(1+\frac{r}{r_0})}\exp\left(\frac{4\,k \pi \arctan(\frac{r}{r_0})(1+\frac{r}{r_0})}{r}   \right)- 2Mr+a^2=0,
\end{equation}
on the other hand, the static limit or ergo surface inner and outer ergosurface is given by $g_{tt}=0$, i.e., 
\begin{equation}
r^2\left(1+\frac{r^2}{r_0^2}\right)^{-\frac{2 \,k \pi}{r}(1-\frac{r}{r_0})}\left(1+\frac{r}{r_0} \right)^{-\frac{4\, k  \pi}{r}(1+\frac{r}{r_0})}\exp\left(\frac{4\,k \pi \arctan(\frac{r}{r_0})(1+\frac{r}{r_0})}{r}   \right)- 2Mr+a^2 \cos^2\theta=0.
\end{equation}

From Fig 1 we observe that for a given $\rho_0$ and $r_0$, one gets two horizons if $a<a_E$. However when $a=a_E$ (blue line) the two horizons coincide, in other words we have
an extremal black hole with degenerate horizons.  Beyond this critical value, $a>a_E$, there is no event horizon and the solution corresponds to a naked singularity. Also  we can observe that the effect of dark matter halo is very small, in fact the effect of dark matter results with a smaller value for $g_{tt}$ and $\Delta$ compared to the Kerr vacuum solution.

\begin{figure}[h!]
\includegraphics[width=0.39\textwidth]{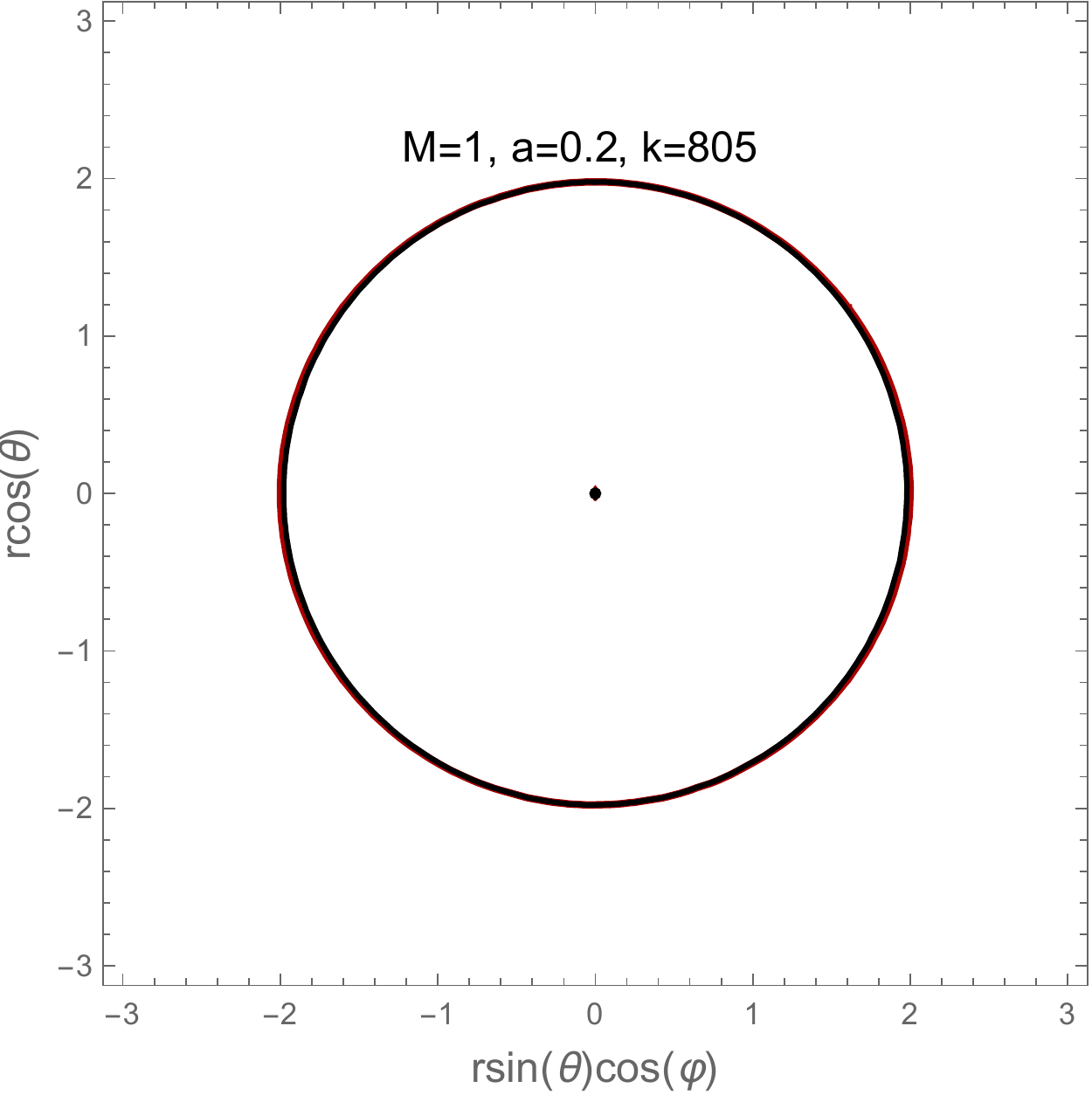}
\includegraphics[width=0.39\textwidth]{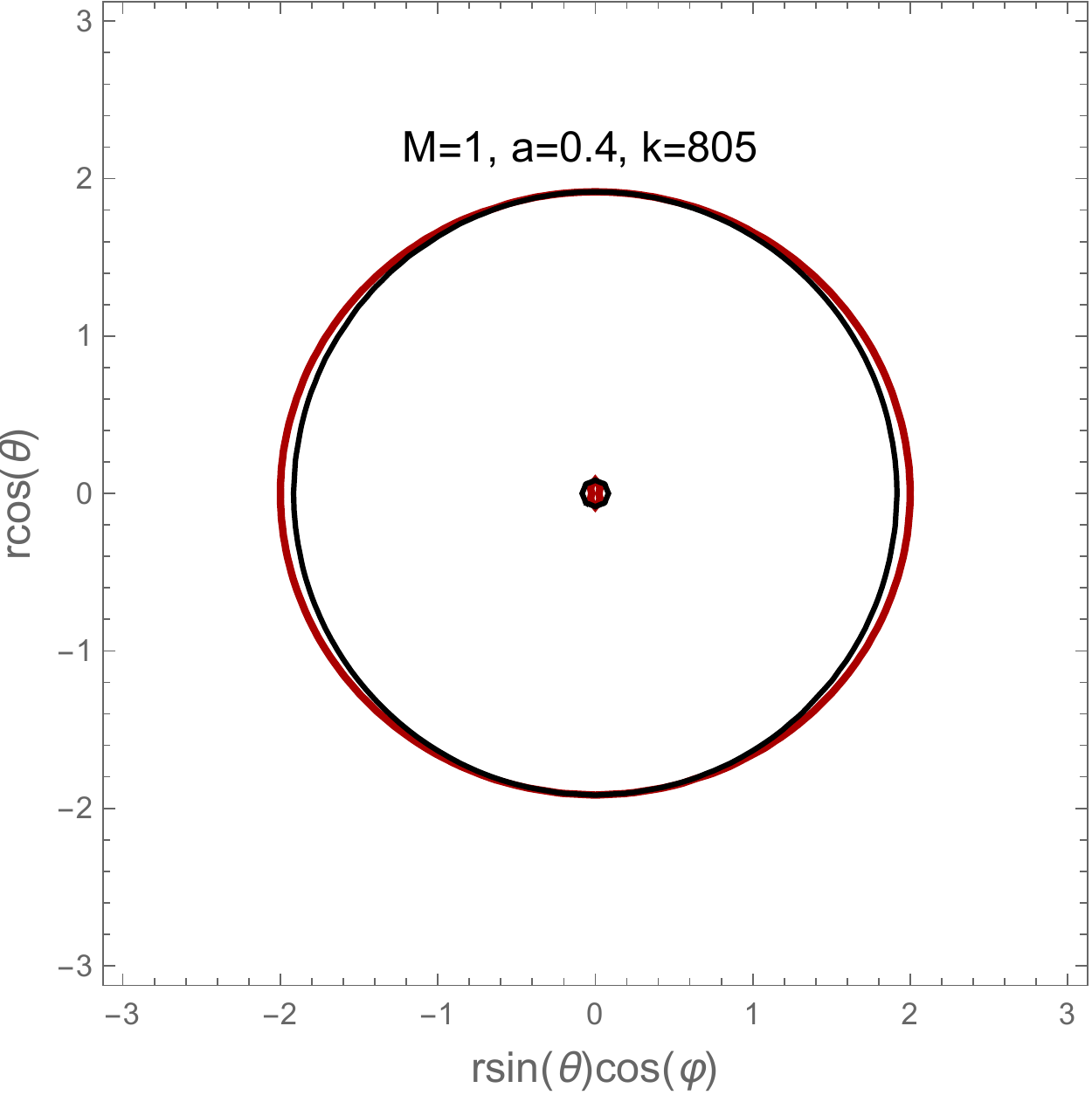}
\includegraphics[width=0.39\textwidth]{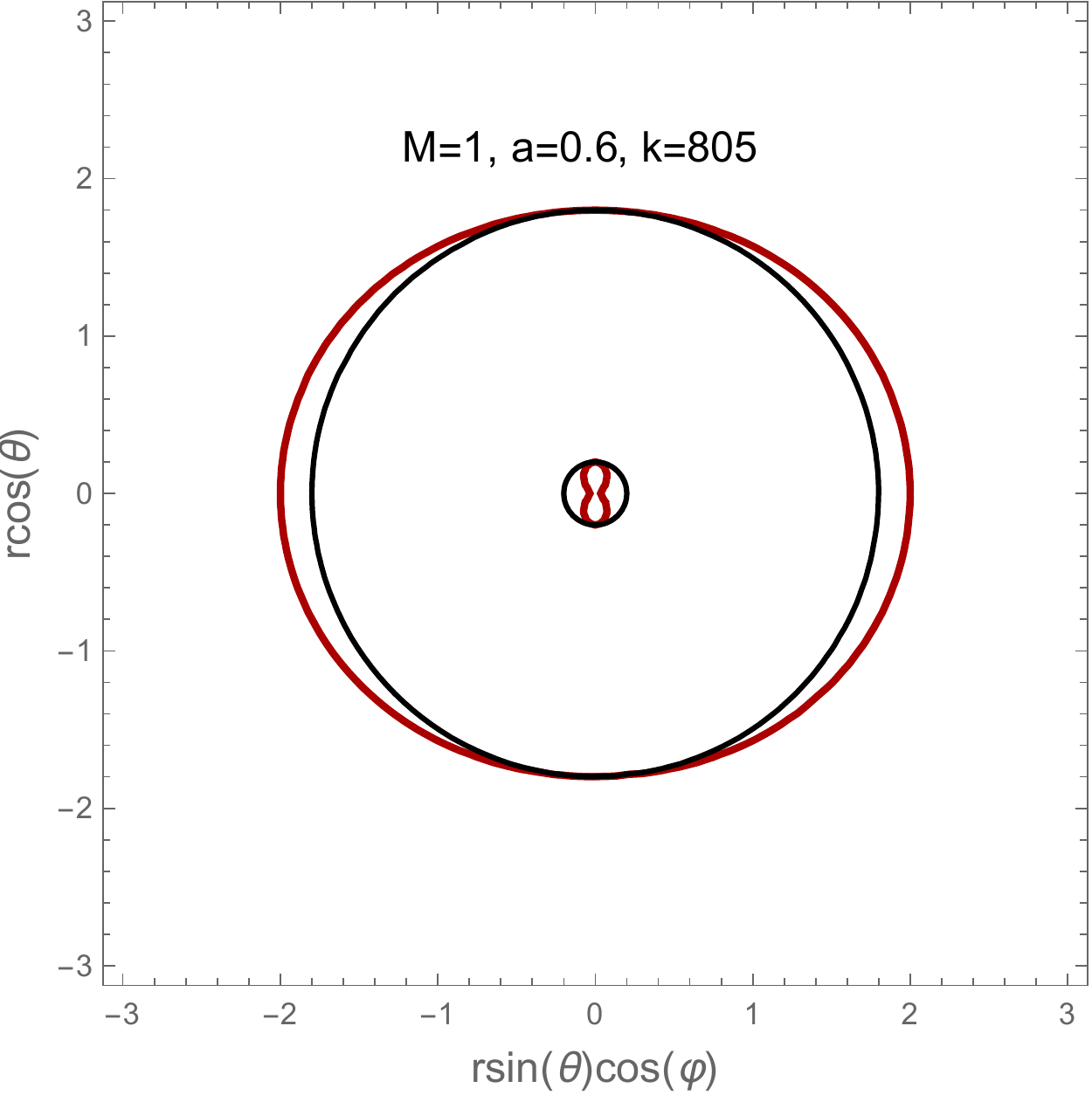}
\includegraphics[width=0.39\textwidth]{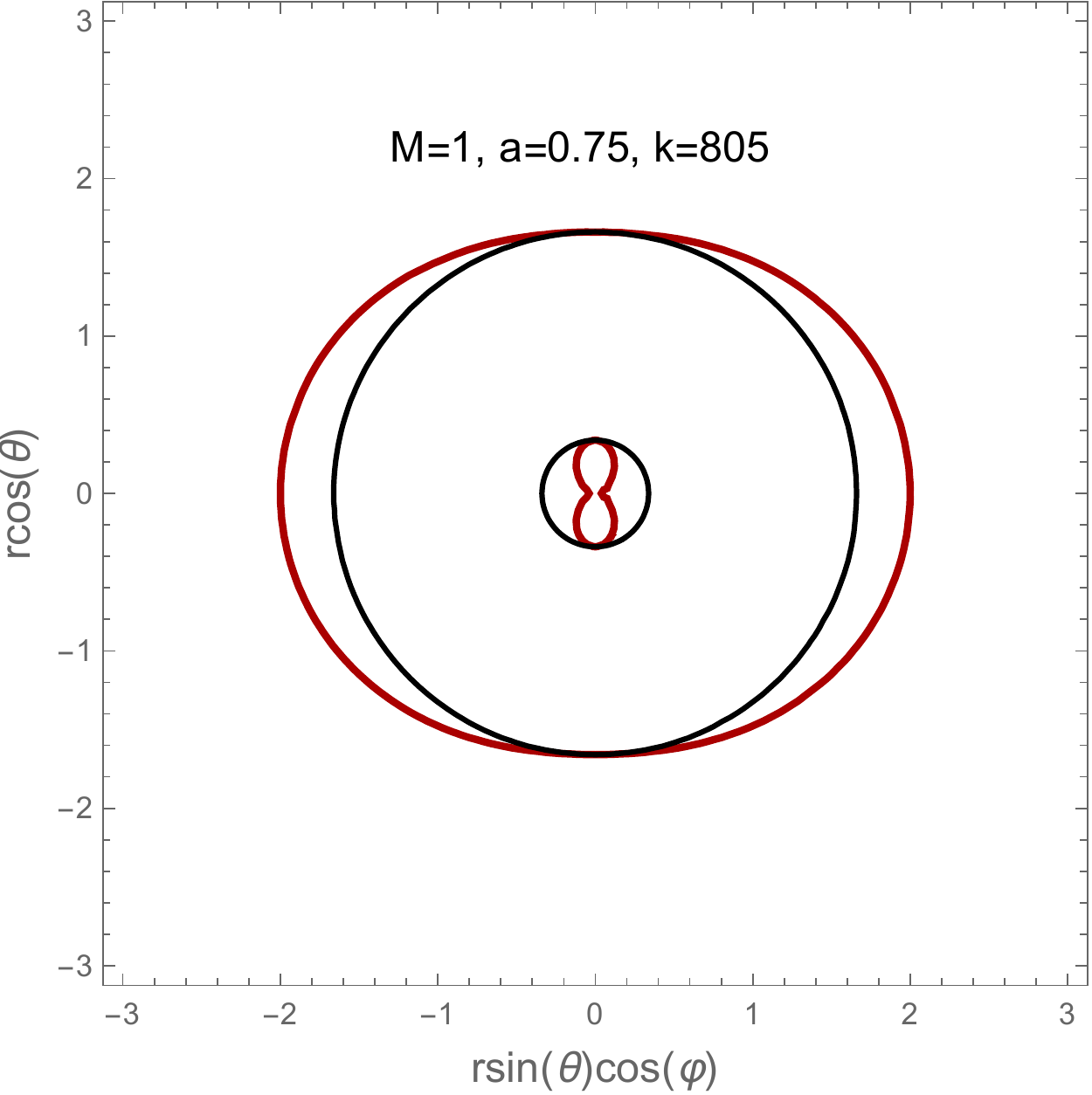}
\includegraphics[width=0.39\textwidth]{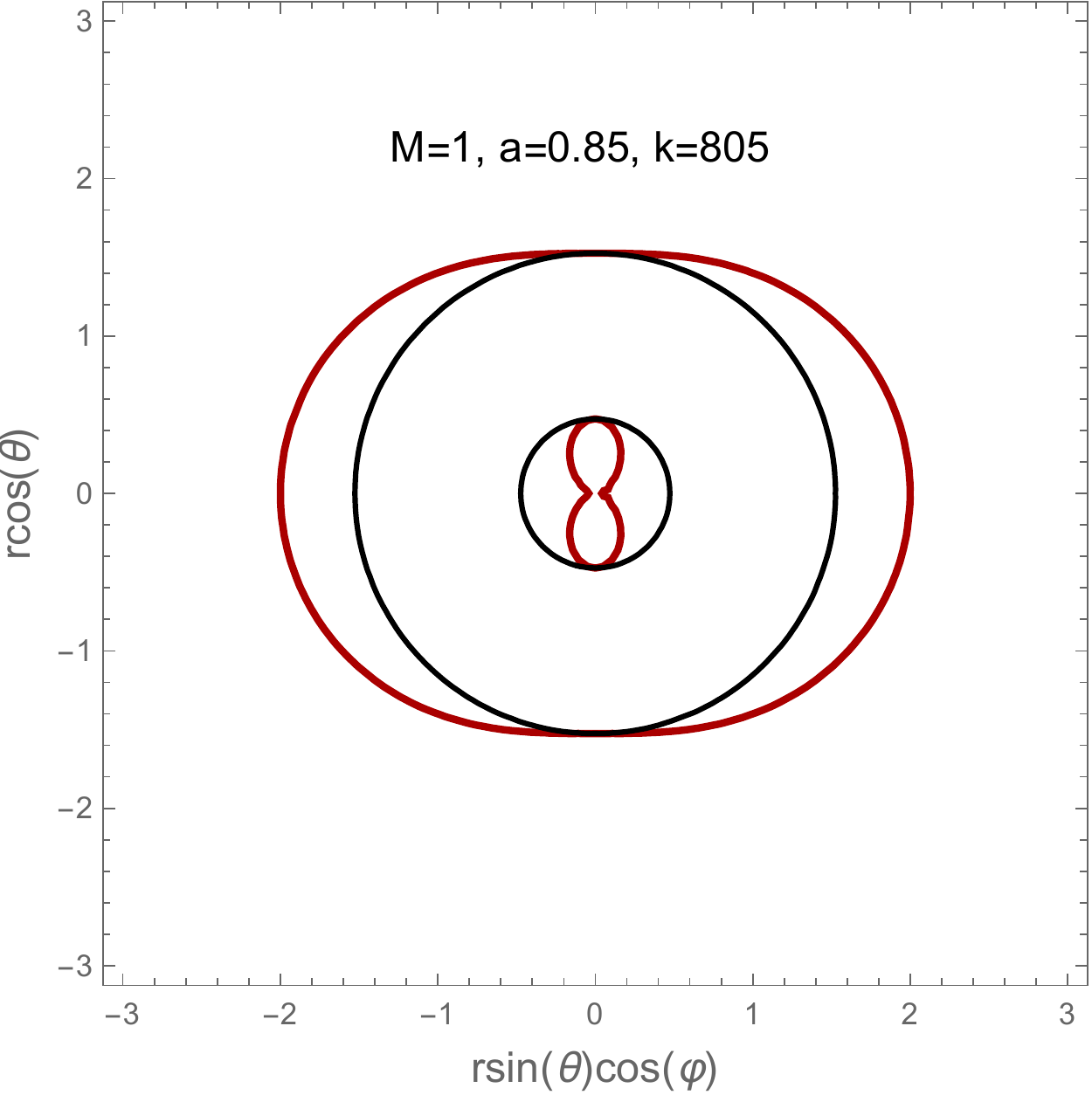}
\includegraphics[width=0.39\textwidth]{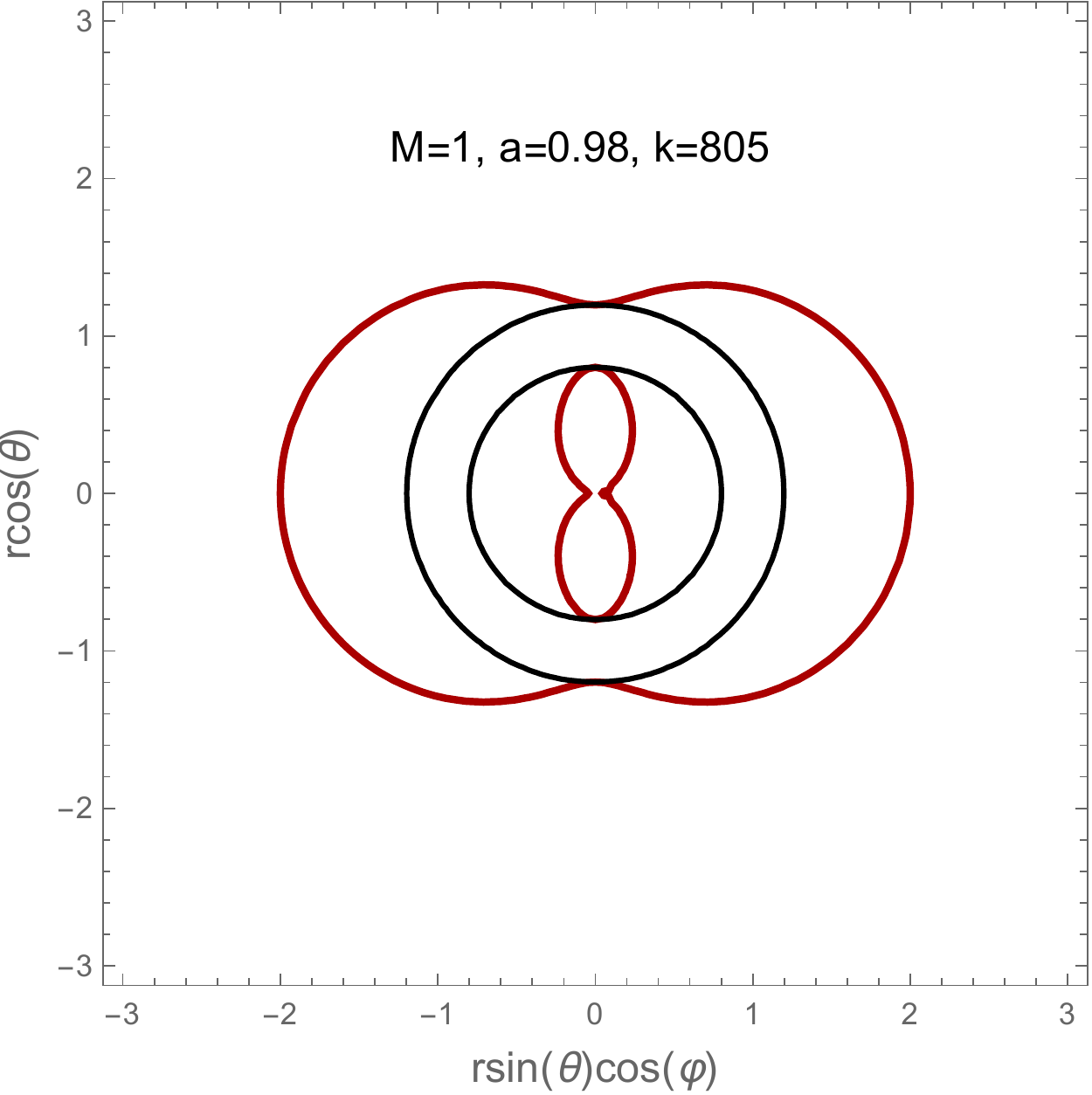}
  \caption{\label{figure0}  Plot showing the variation of the shape of ergoregion  of the rotating black hole with dark matter effects in the $xz$-plane for
different values of $a$. The red line corresponds to the static limit surfaces while the two black lines correspond to the two horizons.}
\end{figure}

\section{Null geodesics}
One of the most important goals of this paper is to study the shadow shapes of the black hole defined by metric \ref{metric}. Towards this goal, we shall firstly analyze the geodesics equations of photons in a given spacetime background. The crucial point behind this method is the fact that one can differentiate the unstable photon orbits which in turn defines the boundary of the shadow.\\
In order to find the  null geodesics around the black hole we can use the Hamilton-Jacobi equation given as follows
\begin{equation}
\partial_\tau\mathcal{J}=-\mathcal{H}.
%\frac{\partial S}{\partial \tau}=-\frac{1}{2}g^{\mu\nu}\frac{\partial S}{\partial {x^{\mu}}}\frac{\partial S}{\partial {x^{\nu}}}
\label{HJ1}
\end{equation}
Note that $\mathcal{J}$ is the Jacobi action, defined in terms of the affine parameter $\tau$ and coordinates $x^\mu$ i.e. $\mathcal{J}=\mathcal{J}(\tau,x^\mu)$ and $\mathcal{H}$ is the Hamiltonian of the particle which can be stated also as $g^{\mu\nu}\partial_\mu\mathcal{J}\;\partial_\nu\mathcal{J}$. From the symmetries of the spacetime, it is well known that along the photon geodesics the energy $E$ and momentum $L$ are conserved quantities, and can be defined by Killing fields $\xi_t=\partial_t$ and $\xi_\phi=\partial_\phi$, respectively. In the case of photon we shall consider also $m=0$. Next, one can separate the Jacobi function as follows
\begin{equation}
\mathcal{J}=\frac{1}{2}m^2\tau-Et+L\phi+\mathcal{J}_r(r)+\mathcal{J}_{\theta}(\theta)
\label{HJ2}
\end{equation}
with $\mathcal{J}_r(r)$ and $\mathcal{J}_\theta(\theta)$ are functions of the coordinates $r$ and $\theta$.
 \begin{figure}[h!]
\includegraphics[width=0.48\textwidth]{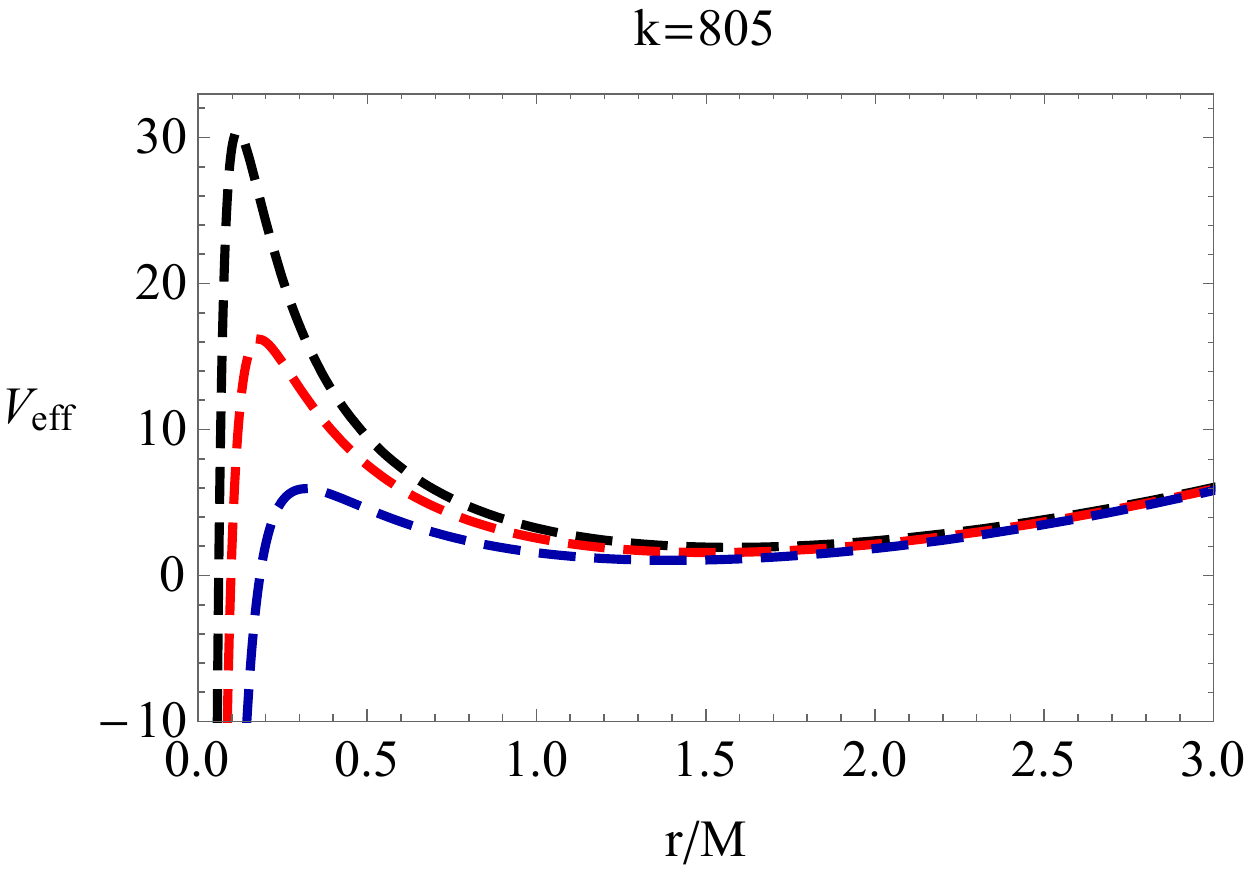}
\includegraphics[width=0.48\textwidth]{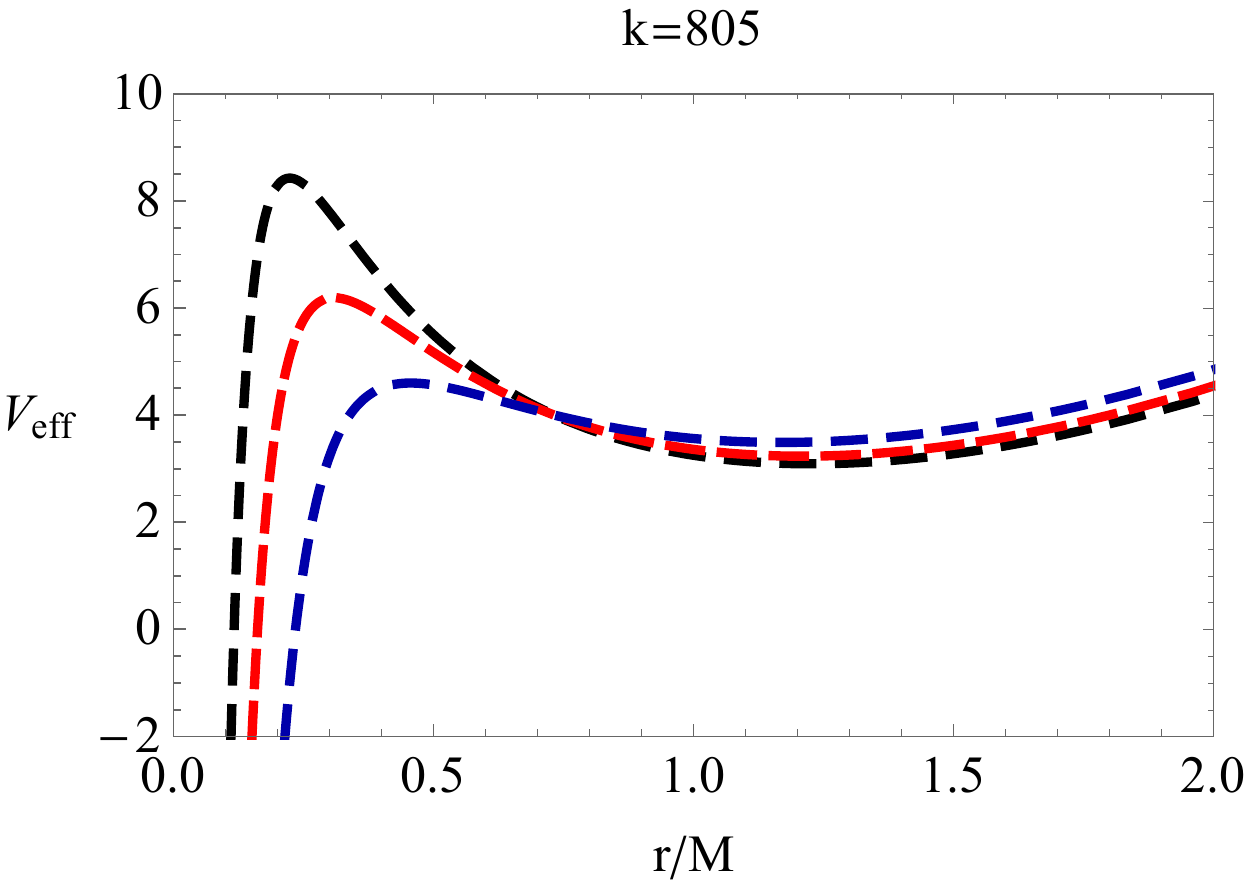}
  \caption{\label{figure0} Left panel: The effective potential of photon moving in equatorial plane, with respect to $r$ in the case of $L \neq 0$. Right panel: The effective potential of photon moving in equatorial plane, with respect to its radial motion with $L=0$. We have chosen $a=0.5$ (black color), $a=0.6$ (red color) and $a=0.75$ (blue color), respectively. }
  %Right panel: We use the same values for $a$ but no dark matter halo effects.  }
  \end{figure}
  
If we combine Eq. (\ref{HJ1}) and Eq. (\ref{HJ2}) it is straightforward to recover the following equations of motions
\begin{align}
\label{HJ3}
&\Sigma\frac{dt}{d\tau}=\frac{r^2+a^2}{\Delta}[E(r^2+a^2)-aL]-a(aE\sin^2\theta-L),\\
&\Sigma\frac{dr}{d\tau}=\sqrt{\mathcal{R}(r)},\\
&\Sigma\frac{d\theta}{d\tau}=\sqrt{\Theta(\theta)},\\
\label{HJ4}
&\Sigma\frac{d\varphi}{d\tau}=\frac{a}{\Delta}[E(r^2+a^2)-aL]-\left(aE-\frac{L}{\sin^2\theta}\right),
\end{align}
with $\mathcal{R}(r)$ and $\Theta(\theta)$ given by
\begin{align}
\label{HJ5}
&\mathcal{R}(r)=[E(r^2+a^2)-aL]^2-\Delta[m^2r^2+(aE-L)^2+\mathcal{K}],\\
&\Theta(\theta)=\mathcal{K}-\left(  \dfrac{L^2}{\sin^2\theta}-a^2E^2  \right) \cos^2\theta,
\end{align}
with $\mathcal{K}$ being the Carter constant.

\section{Circular Orbits}
Let us note that the photons which are emitted from the light source can eventually fall into the black hole or scatter away from it. This process defines a region separating these photons forming the contour of the shadow. We can analyze the presence of unstable circular orbits around the black hole be writing the radial geodesic equation in terms of effective potential $V_{\text{eff}}$ corresponding to the photon's radial motion as
\begin{equation}
\Sigma^2\left(\frac{dr}{d\tau}\right)^2+V_{\text{eff}}=0.
\end{equation}

For our convenience we introduce two independent parameters $\xi$ and $\eta$ \cite{Chandrasekhar:1992} as
\begin{equation}
\xi=L/E,  \;\; \;\; \;\;   \eta=\mathcal{K}/E^2.
\end{equation}
The effective potential in terms of these two parameters is then expressed as
\begin{equation}\label{veff}
V_{\text{eff}}=\Delta((a-\xi)^2+\eta)-(r^2+a^2-a\;\xi)^2,
\end{equation}
where we have replaced $V_{\text{eff}}/E^2$ by $V_{\text{eff}}$. Figure (\ref{figure0}) shows the variation in effective potential associated with the radial motion of photons. Now the circular photon orbits exists when at some  constant $r=r_c$ the conditions
\begin{equation}\label{cond}
V_{\text{eff}}(r)=0,\quad~~~\frac{dV_{\text{eff}}(r)}{dr}=0
\end{equation}
are satisfied. We then use these equations to obtain the following results
\begin{eqnarray}
\xi&=&\frac{(3Mr^2-a^2(M+2r))\mathcal{A}+\exp(\frac{4 k \pi (r+r_0)\arctan(\frac{r}{r_0})}{r r_0} (r (a^2-r^2)+ k\pi (a^2+r^2)\mathcal{B})}{a \exp(\frac{4 k \pi (r+r_0) \arctan(\frac{r}{r_0})}{r r_0})(r+ k \pi \mathcal{B})-a M \mathcal{A} }\\\notag
\eta&=&\frac{ r^3\left( M( 4 a^2-9Mr) \mathcal{A}^2-r\exp(\frac{8 k \pi (r+r_0) \arctan(\frac{r}{r_0}}{r r_0})(r+2 k \pi \arctan(\frac{r}{r_0})-k \pi \mathcal{K})^2+2 \mathcal{A}\exp(\frac{4 k \pi (r+r_0)\arctan(\frac{r}{r_0})}{r r_0})\mathcal{N}\right)}{\left(a M \mathcal{A} -a \exp(\frac{4 k \pi (r+r_0) \arctan(\frac{r}{r_0})}{r r_0})(r+ k \pi \mathcal{B})\right)^2}
\end{eqnarray}
where 
\begin{eqnarray}
\mathcal{A}&=&(1+\frac{r^2}{r_0^2})^{\frac{2 k \pi (r_0-r)}{r r_0}}(1+\frac{r}{r_0})^{\frac{4 k\pi(r+r_0)}{r r_0}},\\
\mathcal{B}&=&-2 \arctan(\frac{r}{r_0})+\ln(1+\frac{r^2}{r_0^2})+2 \ln (1+\frac{r}{r_0}),\\
\mathcal{K}&=&\ln(1+\frac{r^2}{r_0^2})+2 \ln (1+\frac{r}{r_0}),\\
\mathcal{N}&=&3 M r^2+k \pi (2 a^2-3 Mr) \mathcal{B}.
\end{eqnarray}

In the special case $k\to 0$ we recover the Kerr vacuum case \cite{hioki}
\begin{align}
\xi=\frac{r^2(3M-r)-a^2(M-r)}{a (r-M)},
\end{align}
\begin{align}
\eta=\frac{r^3\left(4 Ma^2 -r (r-3M)^2\right)}{a^2 (r-M)^2}.
\end{align}

Note that $a$ is nonzero in the last two equations. 
\begin{figure}[h!]
  \includegraphics[width=0.47\textwidth]{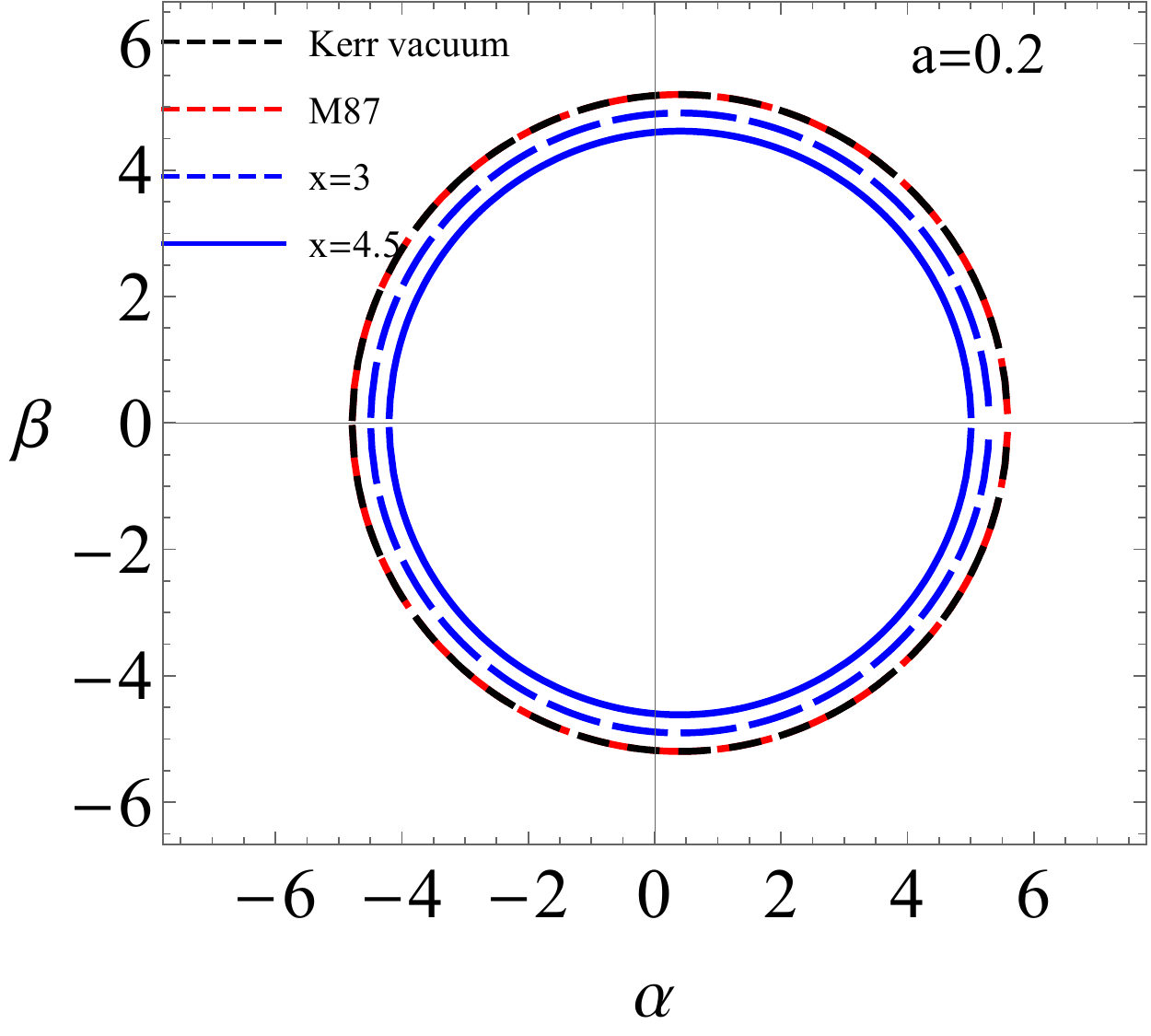}
 \includegraphics[width=0.47\textwidth]{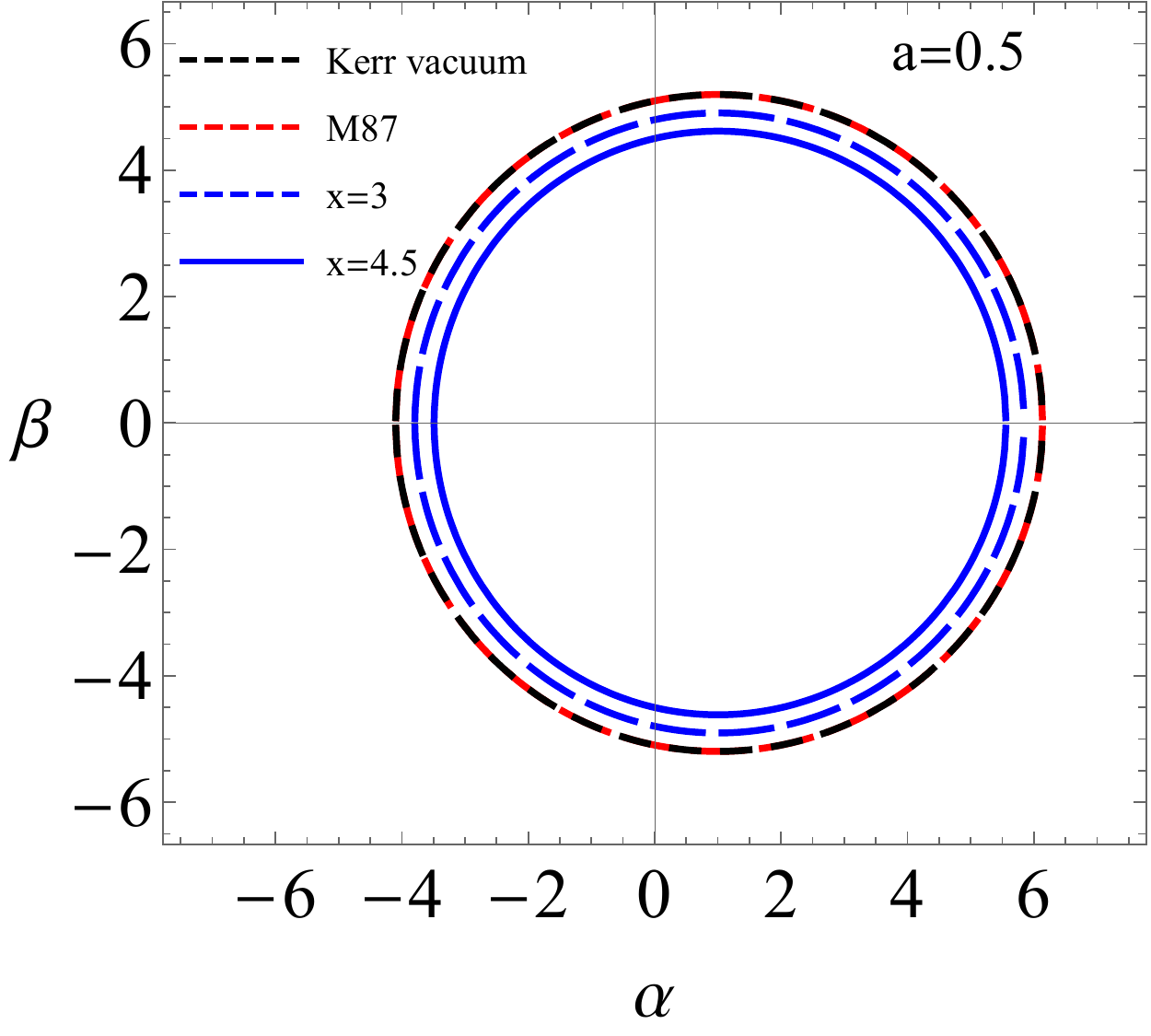}
    \includegraphics[width=0.47\textwidth]{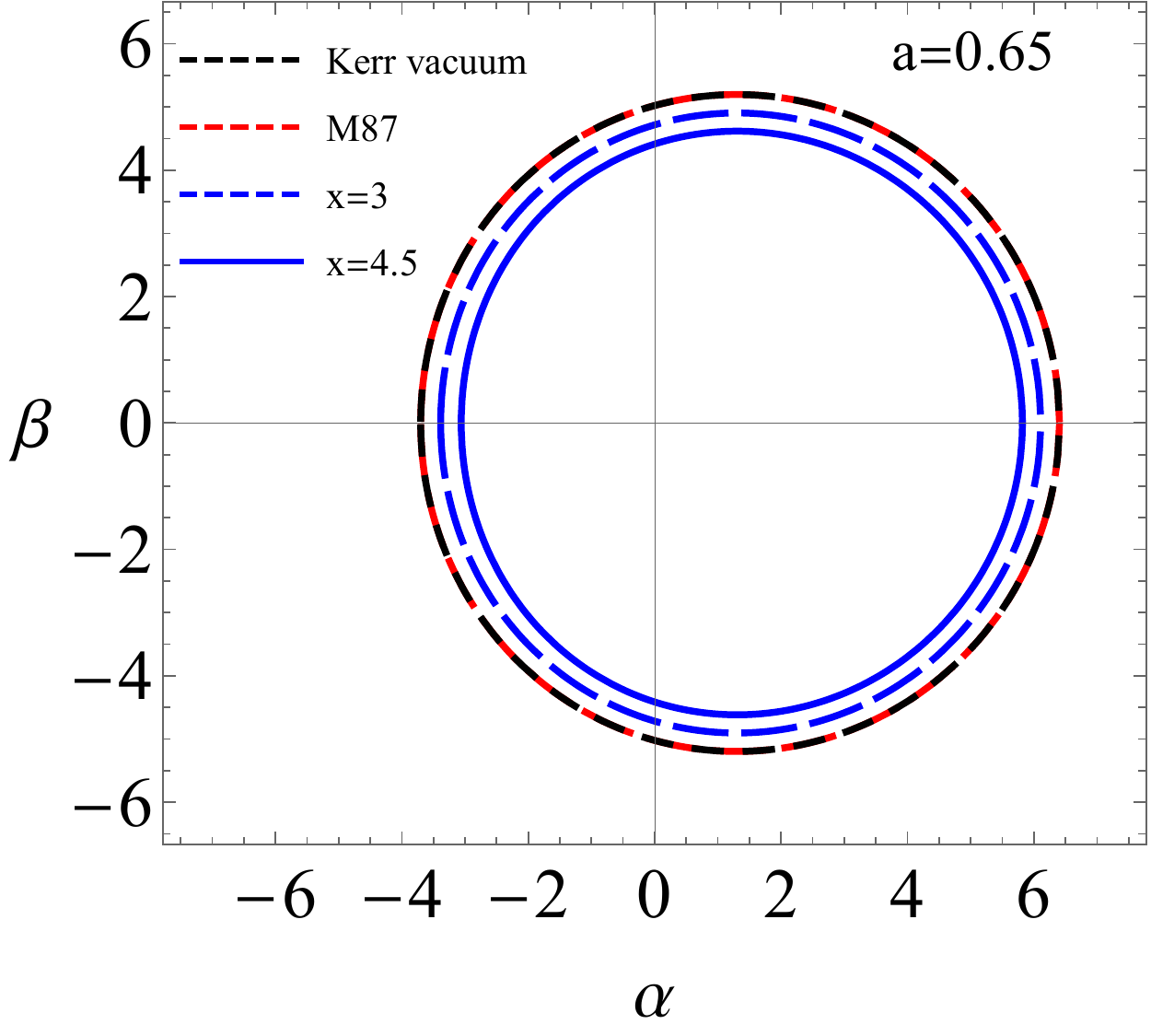}
\includegraphics[width=0.47\textwidth]{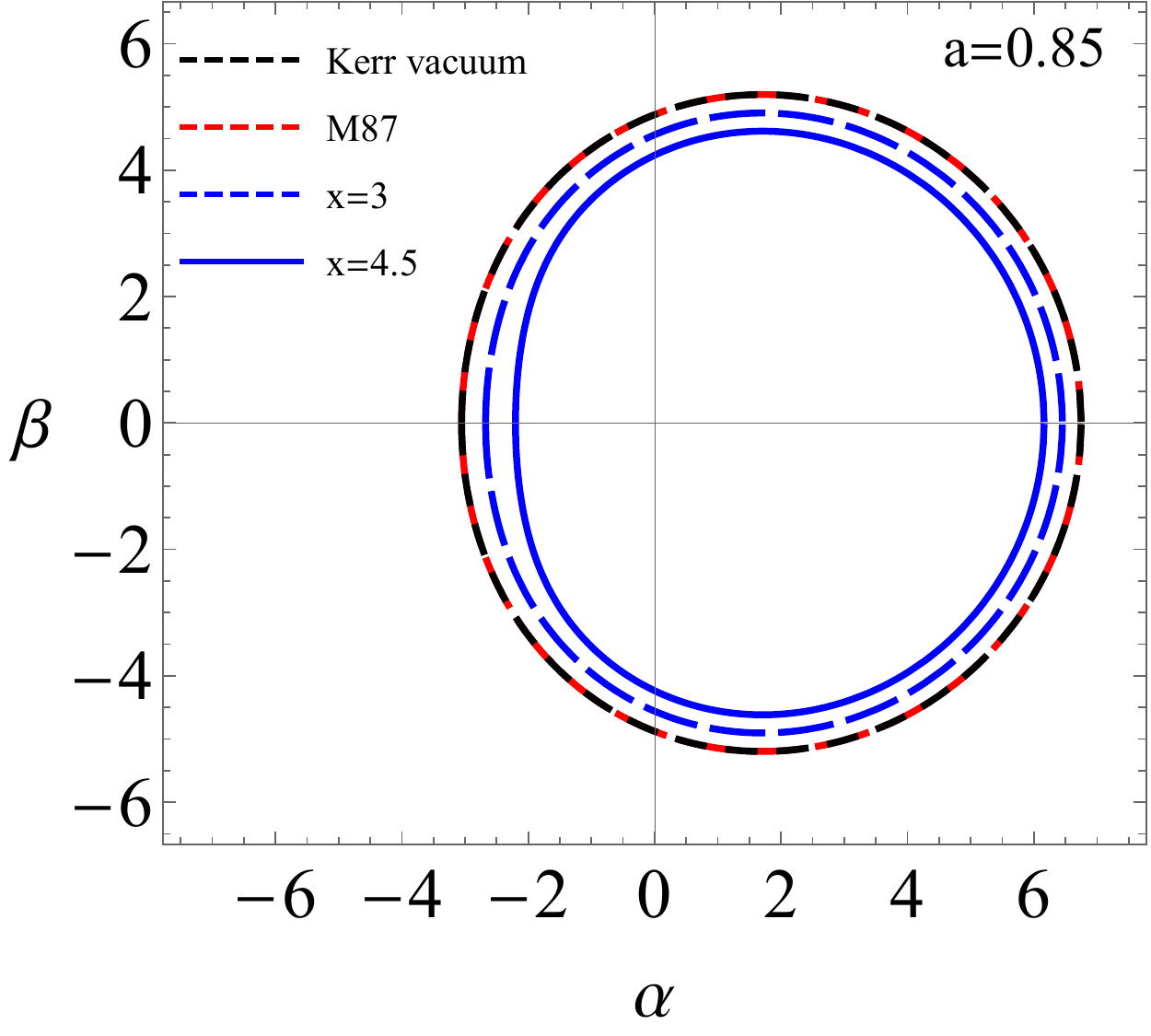}
  \caption{\label{figure1}Variation in shape of shadow for different values of $a$. We use $M=1$ in units of the M87 black hole mass given by $M_{BH}=6.5 \,\times 10^{9} $M\textsubscript{\(\odot\)} and $r_0=91.2$ kpc or $r_0=28.8\,\times 10^7 M_{BH}$. Furthermore for M87 we have used $\rho_0=6.9\,\times 10^{6} $M\textsubscript{\(\odot\)}/kpc$^3$, thus we find $k=805$ in units of $M_{BH}$. The red curve corresponds to M87, black curve corresponds to Kerr vacuum solution and the dotted and line blue curve corresponds to $k=0.005$ with $x=3$ and $x=4.5$, respectively. 
  }
  \end{figure}

\begin{figure}[h!]
  \includegraphics[width=0.47\textwidth]{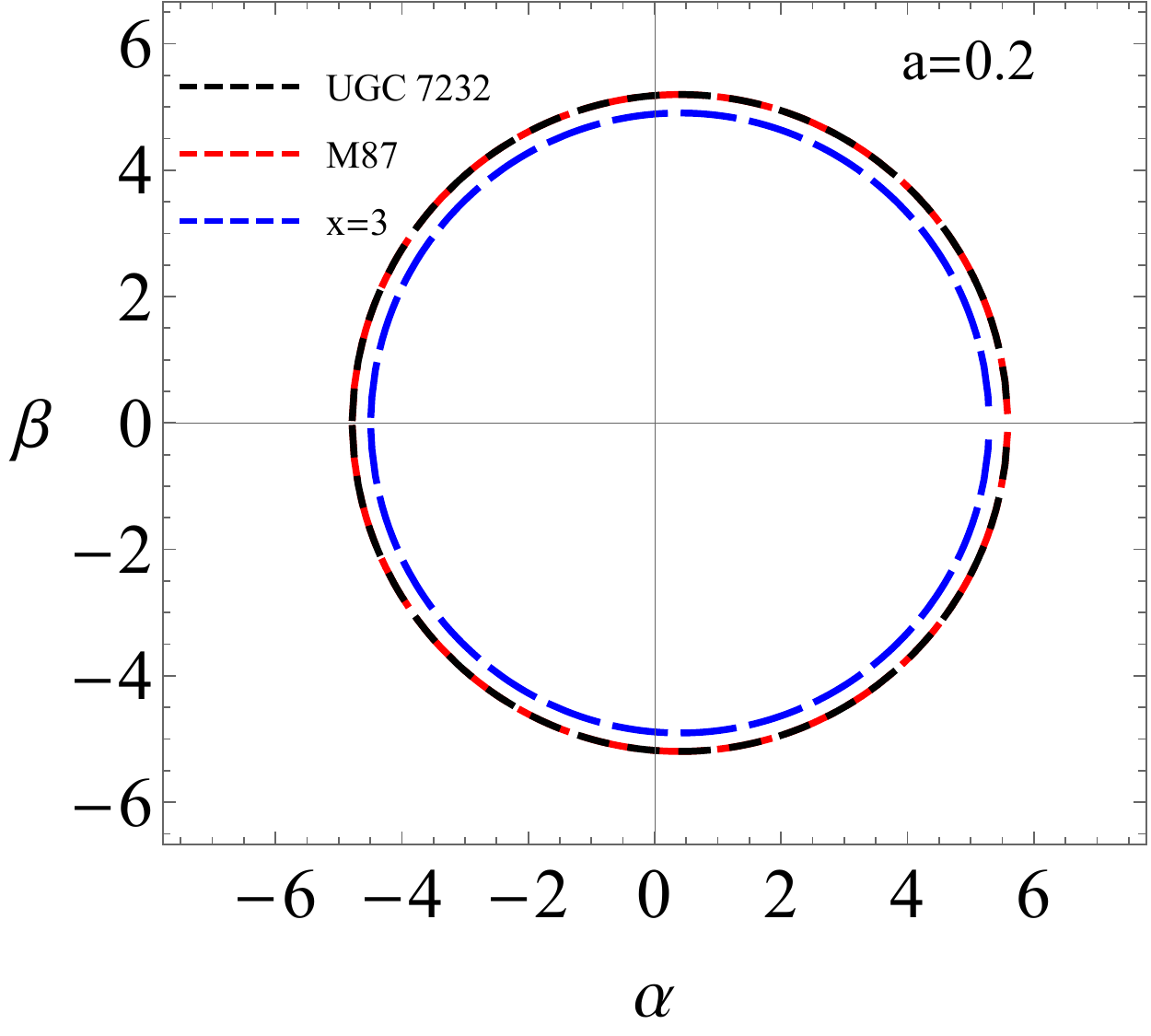}
 \includegraphics[width=0.47\textwidth]{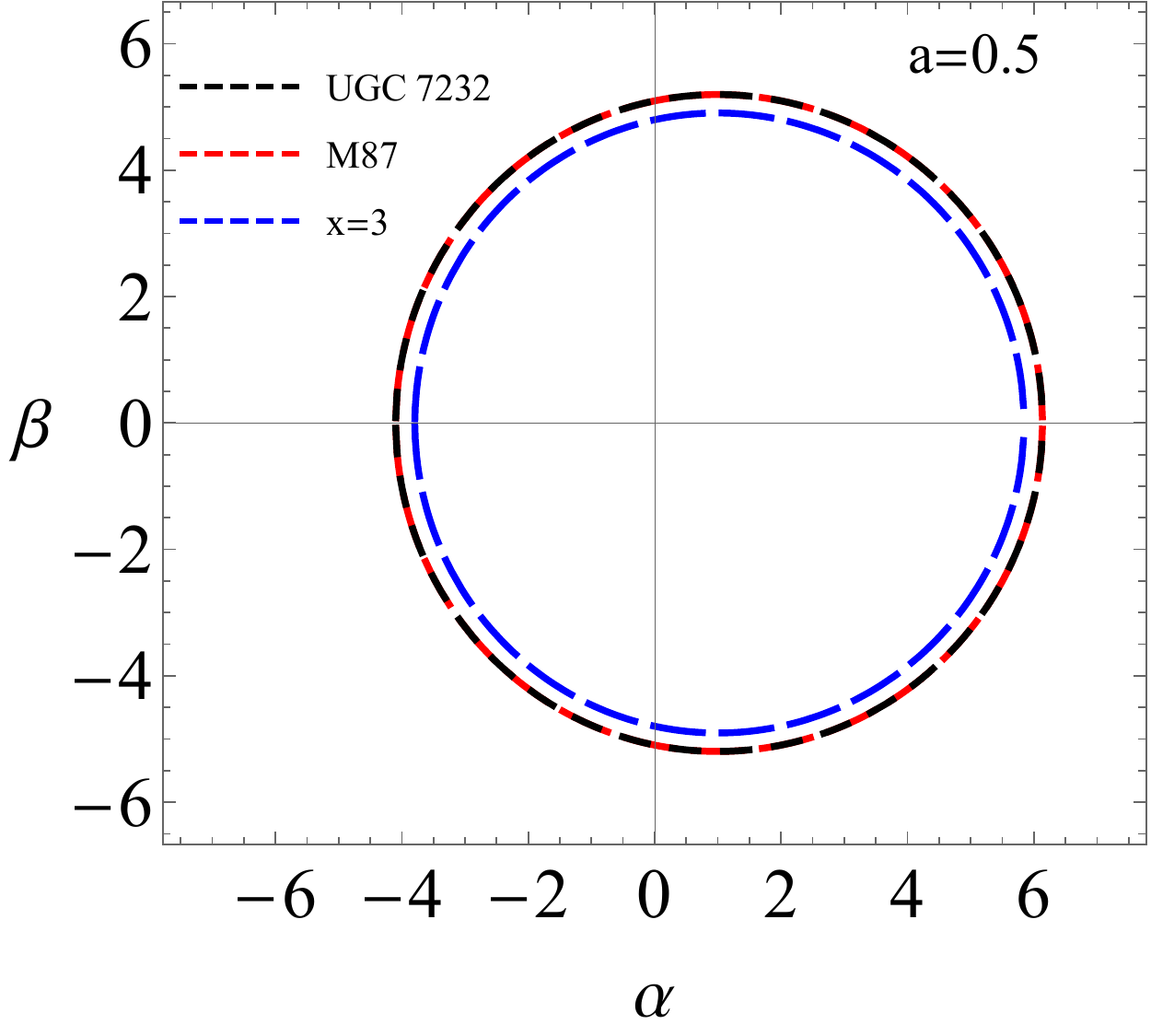}
    \includegraphics[width=0.47\textwidth]{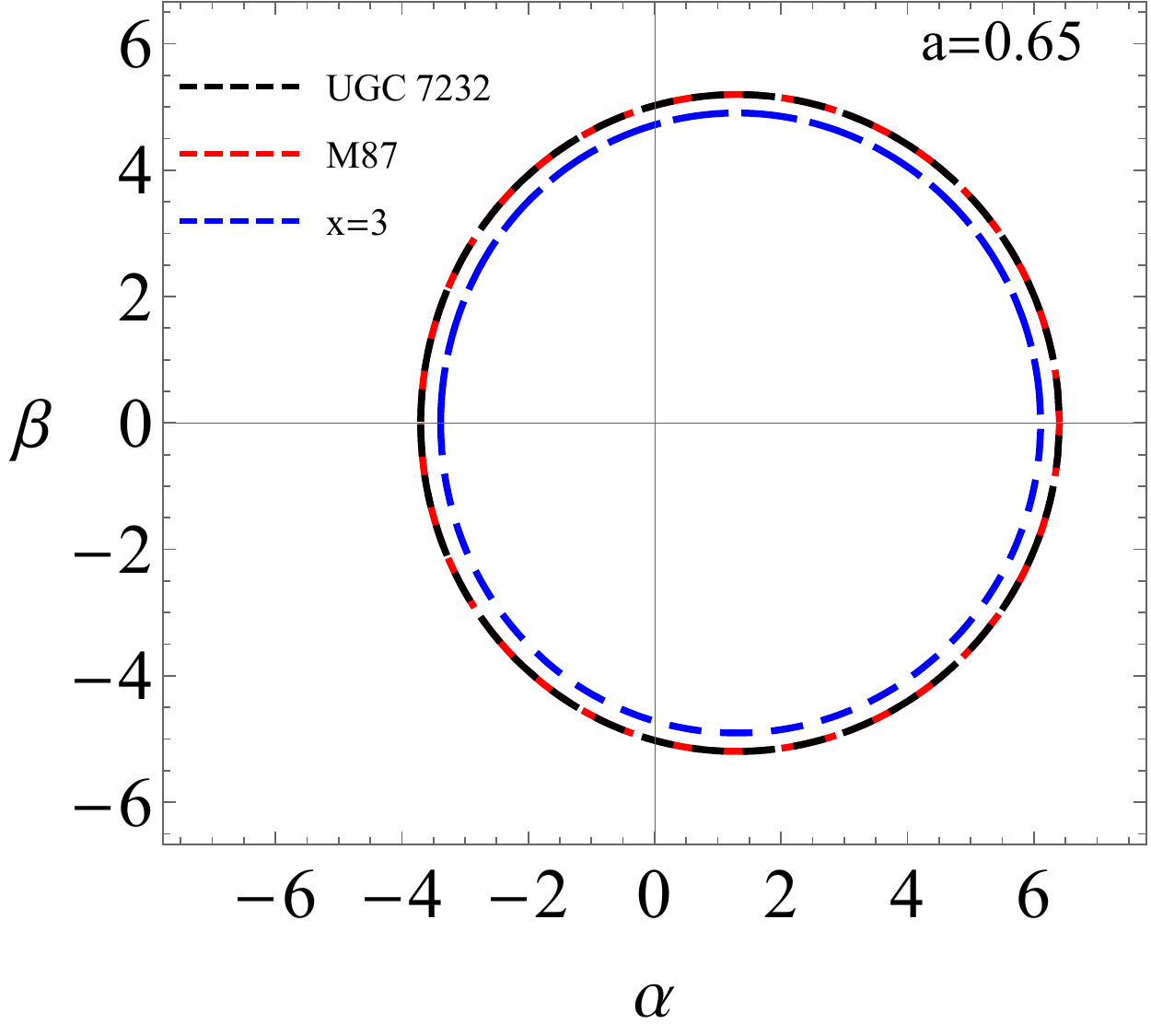}
\includegraphics[width=0.47\textwidth]{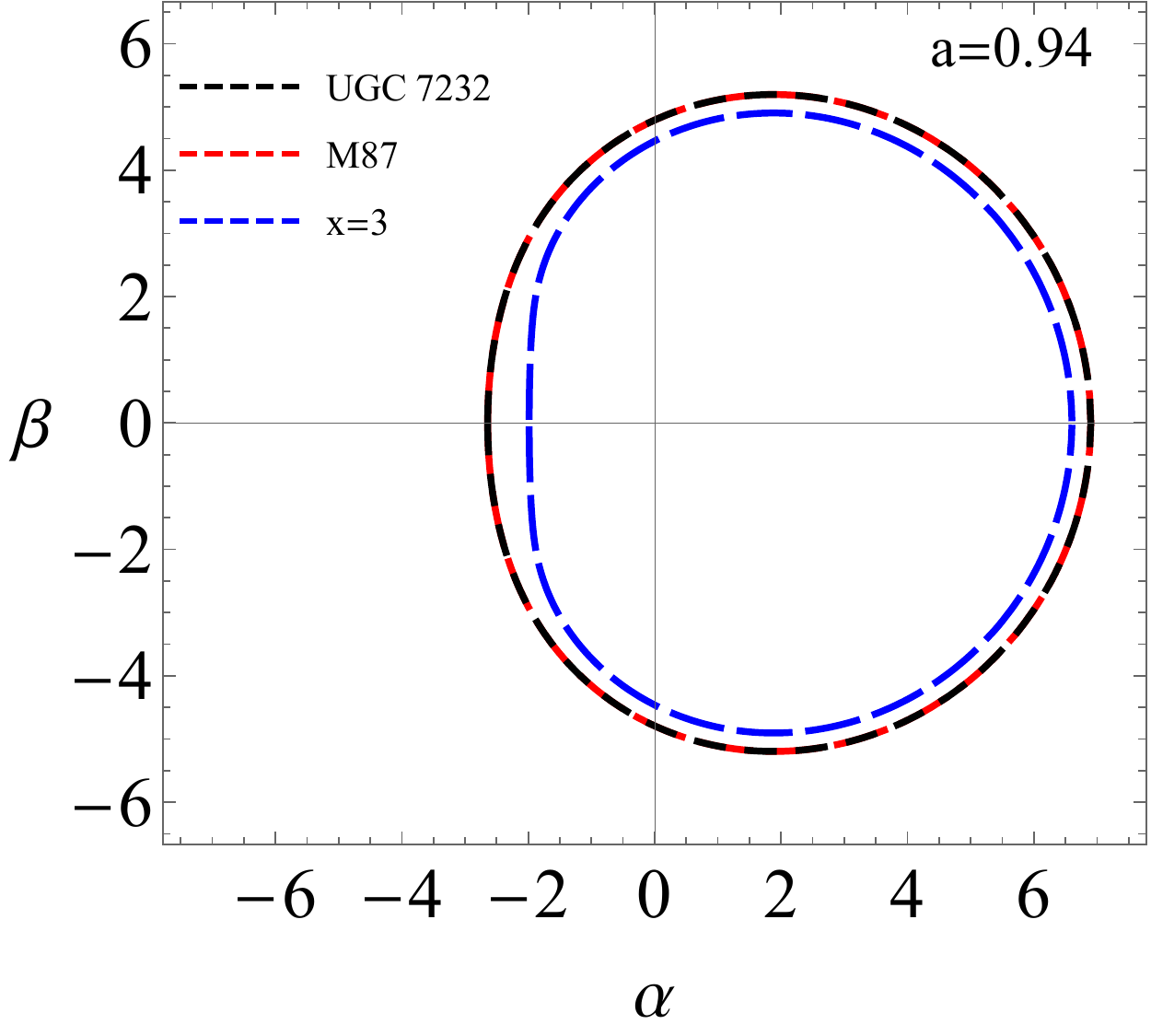}
  \caption{\label{figure1}Variation in shape of shadow for different values of $a$ in the case of small spiral UGC 7232 dominated by dark matter with black hole mass $M_{BH}=10^{5} $M\textsubscript{\(\odot\)}. Here we have used $\rho_0= 7 \times 10^{-23}$ g/cm$^3 $ with a core radius $r_0=0.35$ kpc, or $r_0=0.7 \times 10^{10}$ in units of $M_{BH}$. In a similar way we find $k=438$ in units of $M_{BH}=10^{6} $M\textsubscript{\(\odot\)}. To see the effect better, we compare the results with the Kerr vacuum solution in all plots, respectively. We clearly see that the situation is almost similar to M87, thus we conclude that the effect of dark matter are almost negligible. 
  }
  \end{figure}
  
\begin{figure}[h!]
  \includegraphics[width=0.48\textwidth]{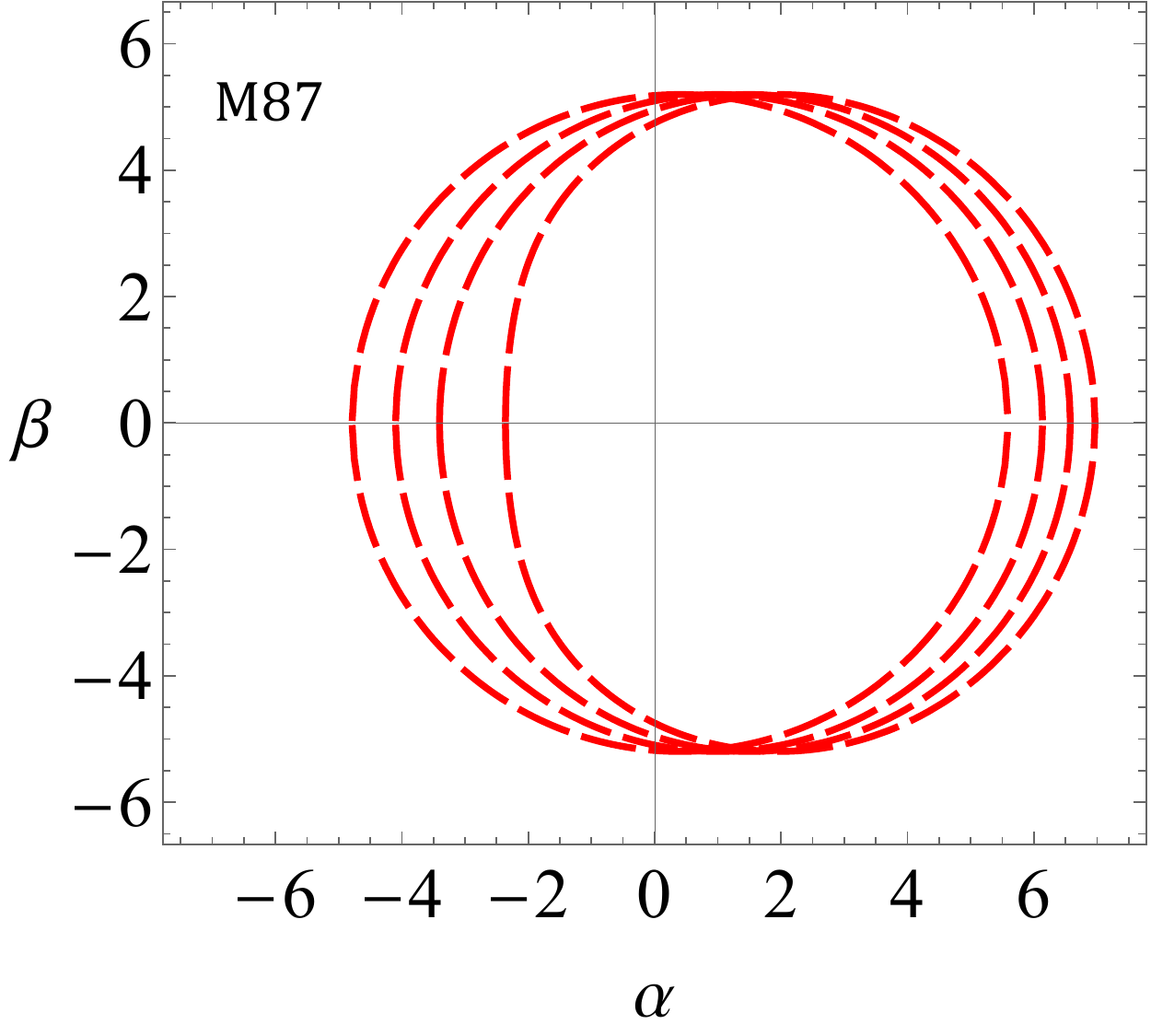}
    \includegraphics[width=0.48\textwidth]{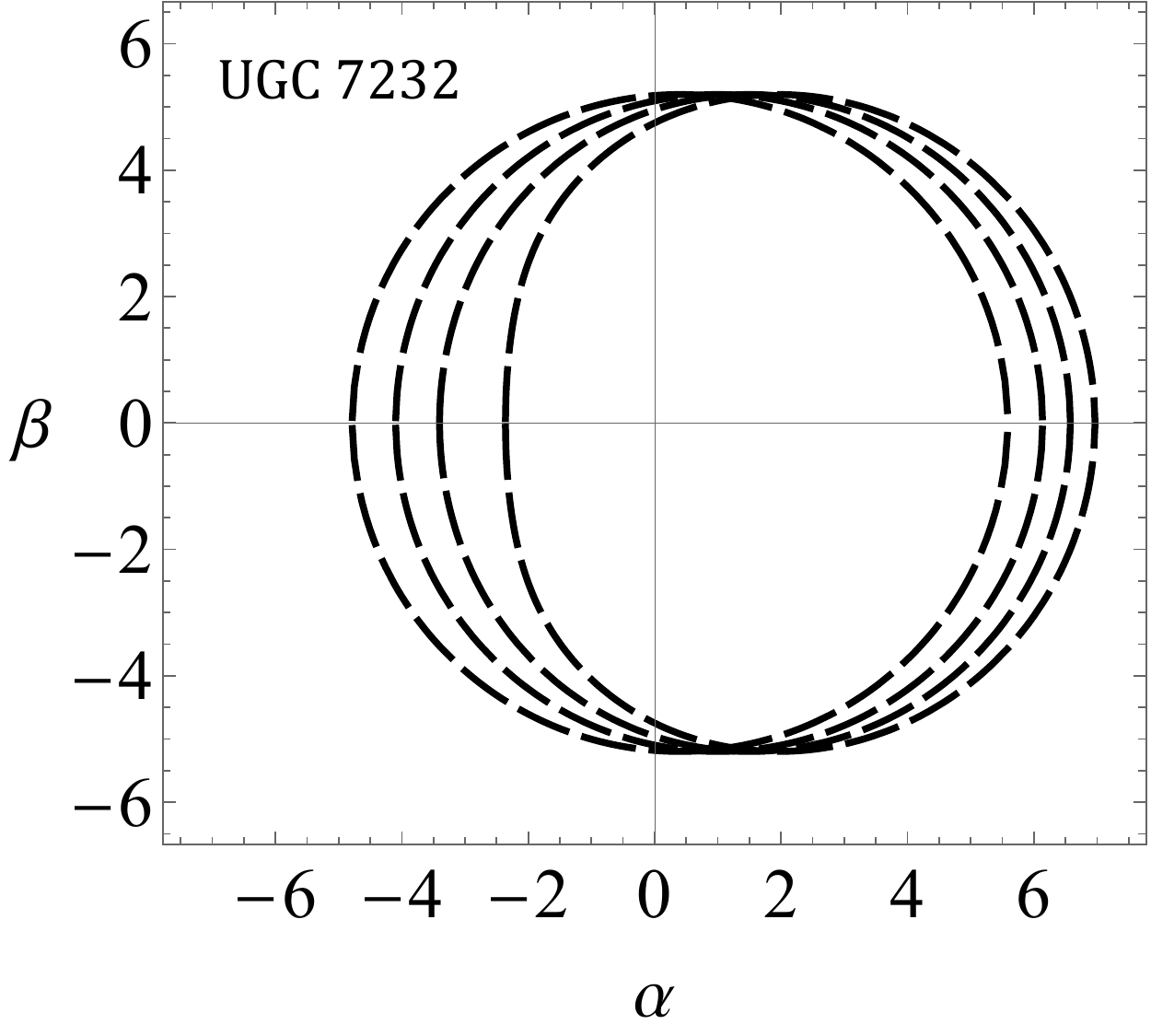}
  \caption{\label{figure1}Left panel: Variation in shape of shadow for M87 for different values of $a$ (dashed lines). Right panel: Variation in shape of shadow in the case of small  spiral UGC 7232 for different values of $a$ (solid lines). We have used $a=(0.2, 0.5, 0.75, 0.98)$ from left to right in both plots. 
  }
  \end{figure}

\newpage
\section{Shadow of M87 black hole surrounded by dark matter}
We can proceed further to find the shadow images related to our our black hole in the presence of dark matter. The observer is located at the position $(r_o,\theta_o)$, where $r_o=r\rightarrow\infty$ and $\theta_o$ being the angular coordinate on observer's sky. Toward this perpouse, we also need to introduce celestial coordinates, say $\alpha$ and $\beta$, given in terms of the following equations  \cite{pfdm}
\begin{align}
& \alpha = \lim_{r_o\to \infty}\left( -r_o^2 \sin \theta_o \dfrac{d\phi}{dr}  \right),\\
& \beta = \lim_{r_o \to \infty}\left( r_o^2 \dfrac{d\theta}{dr}  \right).
\end{align}

Furthermore, one can rewrite these coordinates in terms of two parameters $\xi$ and $\eta$, given by
\begin{align}\nonumber
& \alpha = -\dfrac{\xi}{\sin \theta},\\\label{beta}
& \beta = \pm \sqrt{\eta + a^2\cos^2\theta -\xi^2\cot^2 \theta}.
\end{align}

To simplify the problem further we shall consider that the observer is located in the equatorial plane ($\theta=\pi/2$), $\alpha$ and $\beta$ yielding
\begin{align}
& \alpha = -\xi,\\
& \beta = \pm \sqrt{\eta }.
\end{align}

In order to see better the effect of dark matter let us introduce a new variable, say $x=r/r_0$ resulting with 
\begin{equation}
F(r)=\left(1+x^2\right)^{-\frac{2 \,k \pi}{r}(1-x)}\left(1+x \right)^{-\frac{4\, k  \pi}{r}(1+x)}\exp\left(\frac{4\,k \pi \arctan(x)(1+x)}{r}   \right)- \dfrac{2M}{r}.
\end{equation}

Figures 7 show the deformation in shapes of the black hole shadow with respect to $\rho_0$ and $a$. It is shown that the shadow of the black hole M87 (red curve) in the presence of dark matter is almost indistinguishable compared the the Kerr vacuum (black curve).  The effect is strong if the core radius of the dark matter is of the order of black hole mass, consequently a much denser dark matter medium compared to $\rho_0$ for M87. This will happen if $x$ increases (or the core radius decreases), while keeping constant $k$, as a result the shadow images decreases. 
For an interesting observation, in Figure 8 we show the shadow for small spiral dominated by dark matter, such as the spiral galaxy UGC 7232 with $\rho_0= 7 \times 10^{-23}$ g/cm$^3 $ and  a core radius $r_0=0.35$ kpc and black hole mass $M_{BH}=10^{5} $M\textsubscript{\(\odot\)}. We have found a similar analyses as in the case of M87. Again under specific condition, say with a very small core radius such as the case  $x=1,2,3,4$ and say $k=0.005$. We find that the shadow shapes are considerable decreased compared to the Kerr vacuum solution. This can be seen from the blue curve in Figs. 7 and 8. 

\subsection{A comparison with NFW profile and the cusp phenomenon}
In Ref. \cite{Hou:2018bar} authors have studied the effect of Cold Dark Matter on the black hole shadow using the well known Navarro-Frenk-White (NFW) dark matter density profile given by
\begin{equation}
\rho_{NFW}=\frac{\rho_0}{\frac{r}{r_s}\left(1+ \frac{r}{r_s}\right)^2}.
\end{equation}
  
In particular they obtained a black hole solution surrounded by dark matter given by the line element
\begin{equation}
F(r)=\left(1+\frac{r}{r_s} \right)^{-\frac{8\, k  \pi}{r}}- \dfrac{2M}{r}.
\end{equation}
with $k= \rho_0 r_s^3$. In the case of M87 we have $r_s=130$ kpc and $\rho_0=0.008\,\times 10^{7.5} $M\textsubscript{\(\odot\)}/kpc$^3$ (see, \cite{auger}). In what follows we shall compare the shadow images using the URC and NFW profile. Fig. 10 show that in the case of M87 shadow images are in perfect agreement, however in a stronger dark matter medium shadow images show different effect. In particular the NFW effect increases the size of shadow images compared to the Kerr vacuum, while the URC profile results in a decrease of shadow images. In order to have strong effect we need a core radius comparable to the black hole mass, and this difference may be related to the “cusp” phenomenon, namely when the distance $r$ from the black hole is below $1-2$ kpc it is known that energy density of dark matter diverges for NFW profile.   
 \begin{figure}[h!]
  \includegraphics[width=0.49\textwidth]{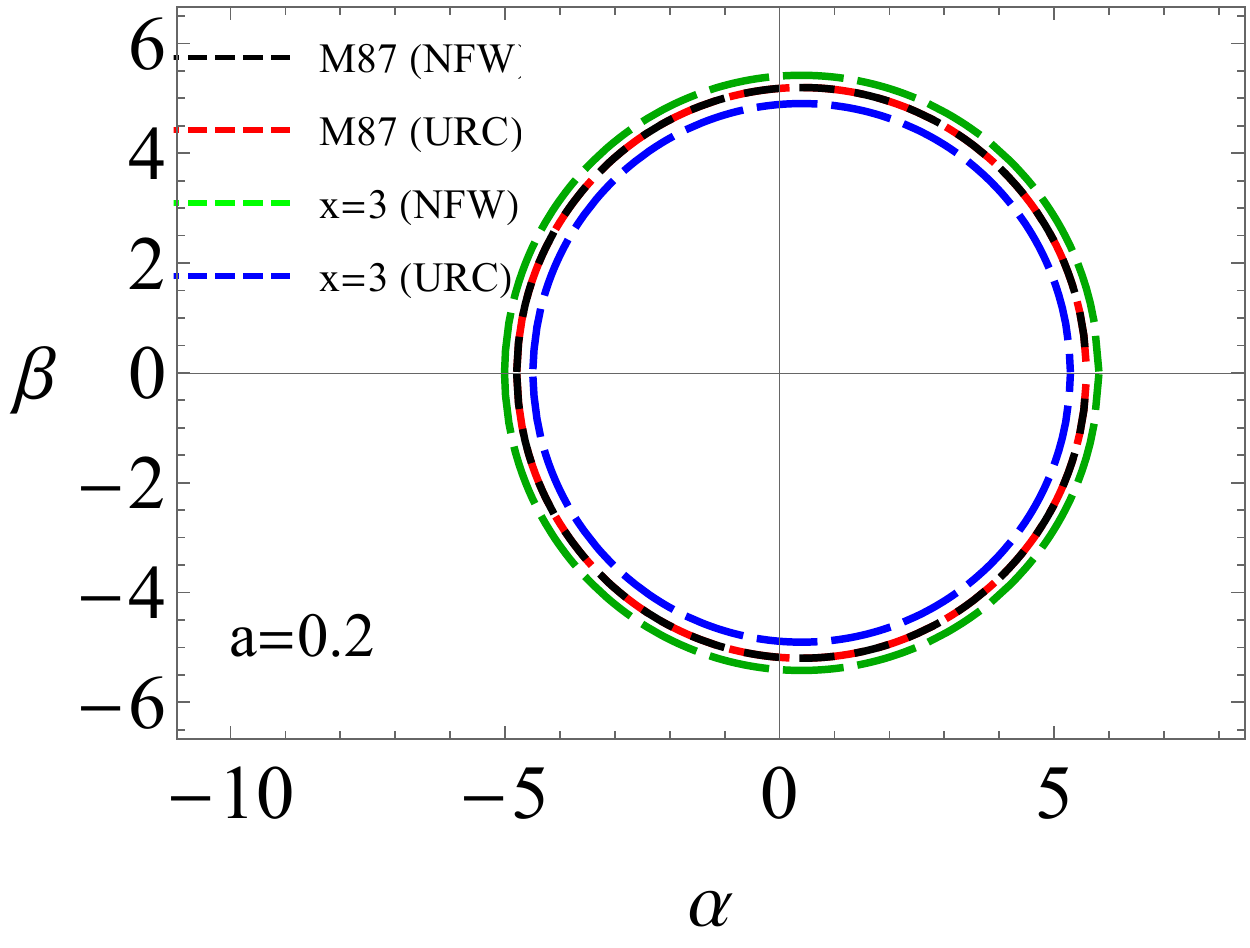}
    \includegraphics[width=0.49\textwidth]{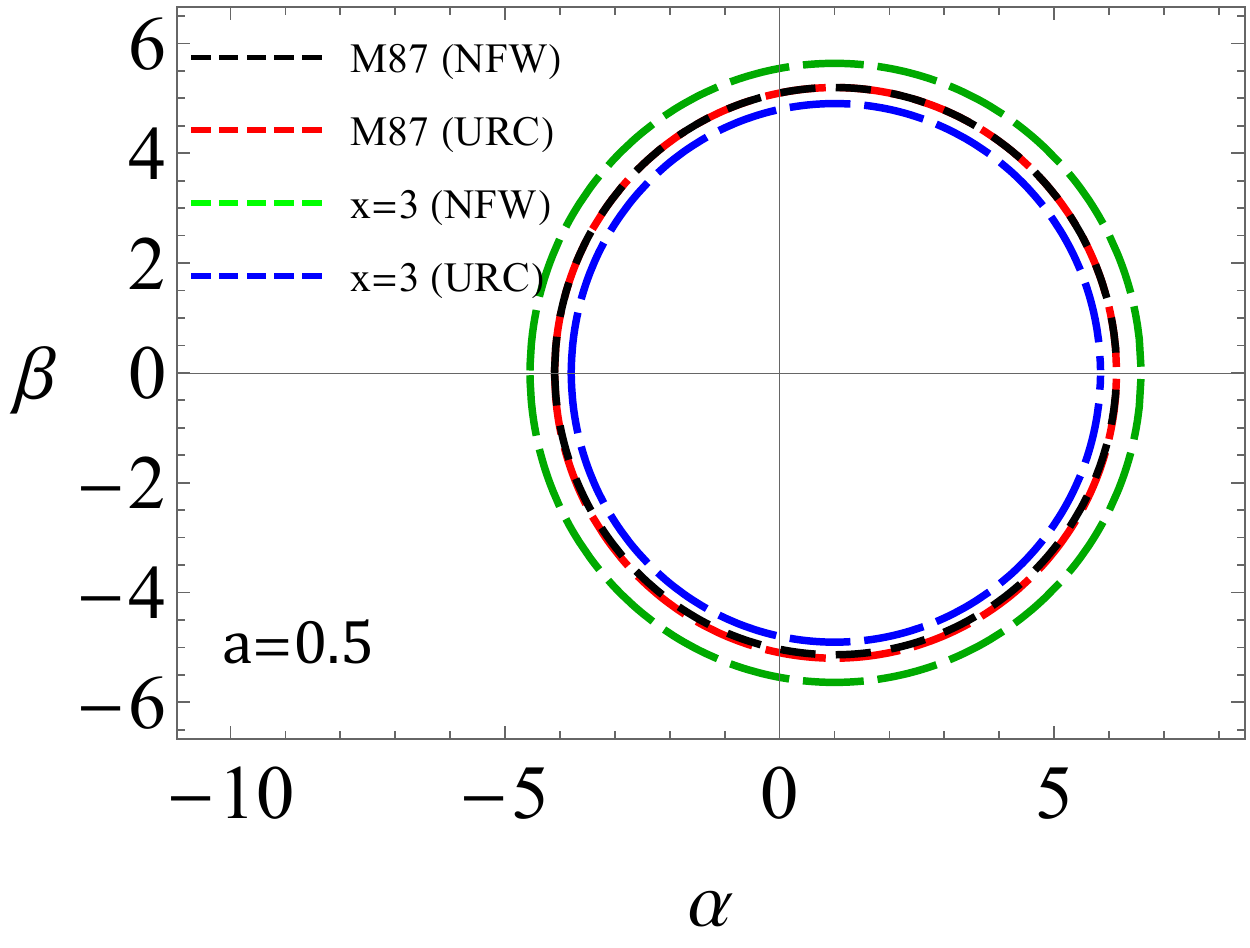}
    \includegraphics[width=0.49\textwidth]{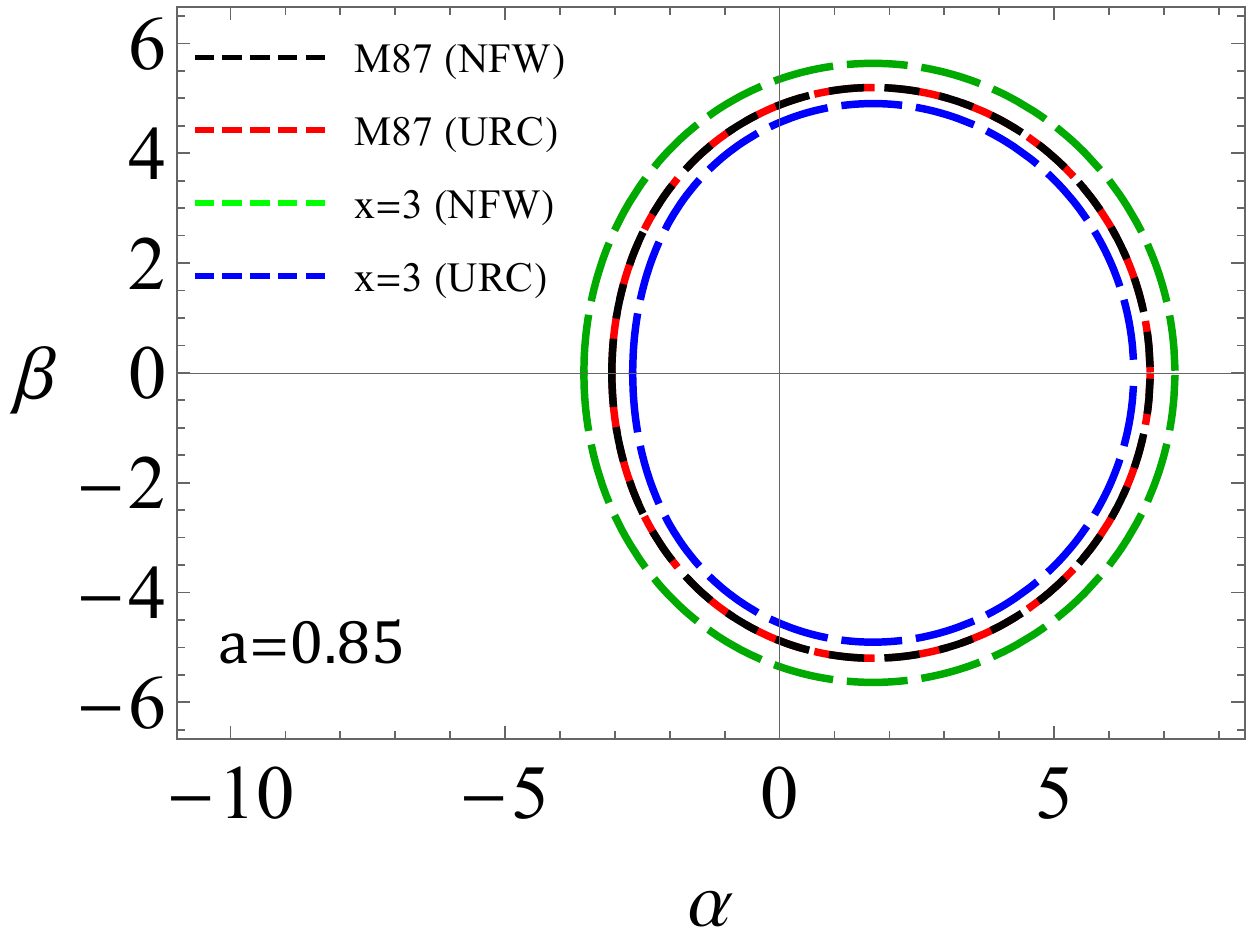}
    \includegraphics[width=0.49\textwidth]{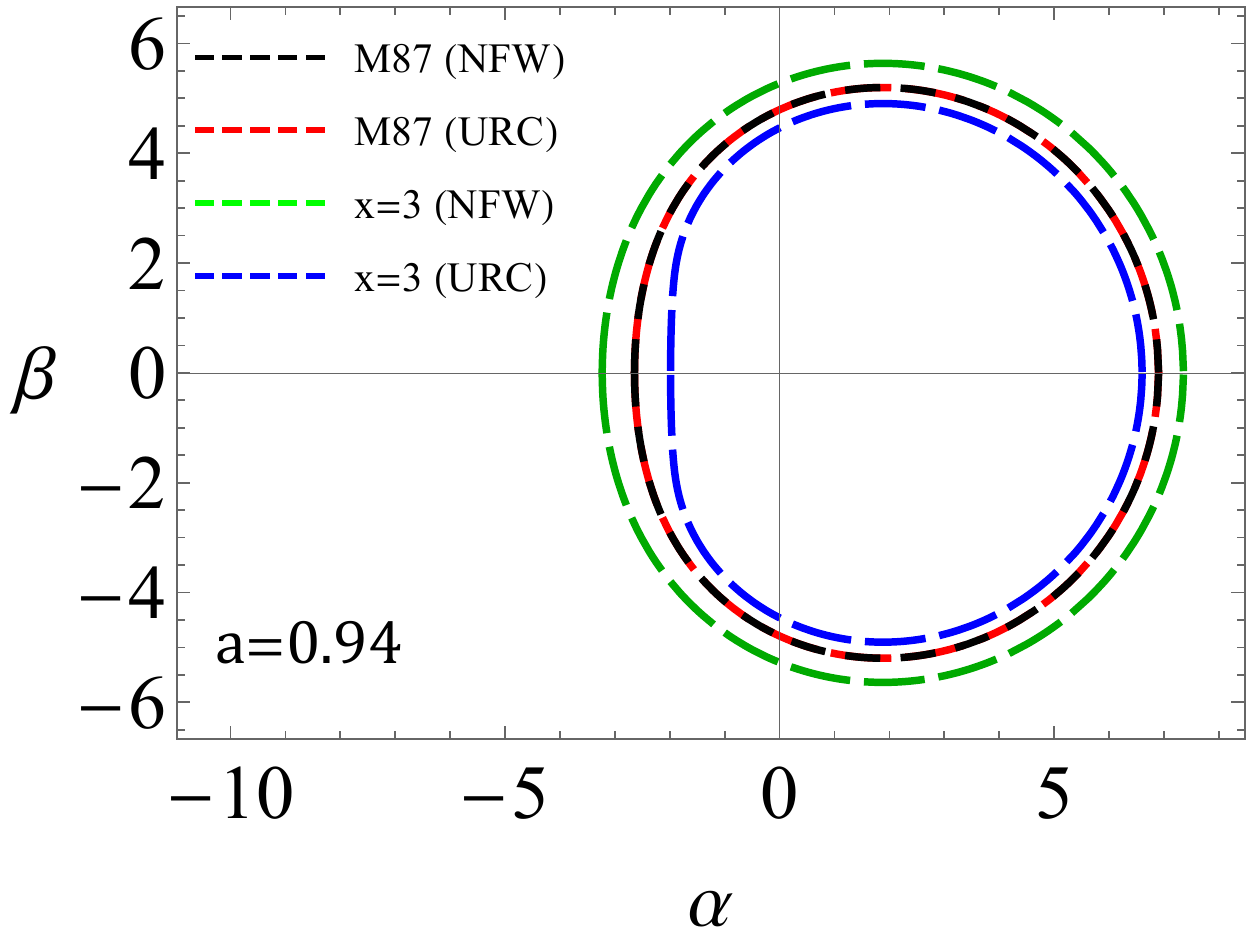}
  \caption{\label{figure1} Variation of shadow shapes of M87 using NFW and URC profile. For the case of M87 the shadow plots are almost same, while for stronger effect they show opposite behavior, namely it is observed that NFW increases the shadow size, on the hand URC decreases the shadow size. We have introduced a new variable $x=r/r_s$ or $x=r/r_0$, respectively.
  }
  \end{figure}

  \begin{figure}[h!]
  \includegraphics[width=0.48\textwidth]{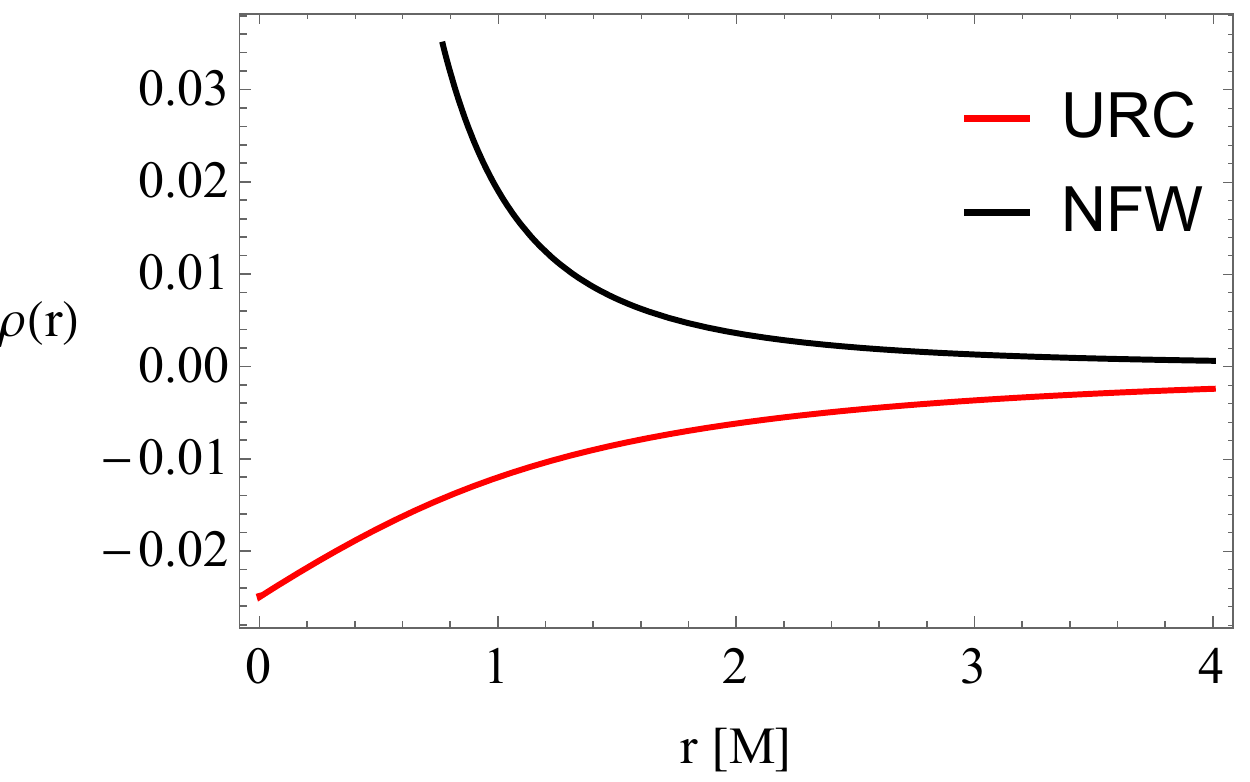}
  \caption{\label{figure1} We show the dark matter for the NFW and URC density profile, respectively. In the case of URC profile the energy density of dark matter is finite.}
  \end{figure}

The energy density of the surrounded dark matter is given by the following equation \cite{Hou:2018bar}
\begin{equation}
\kappa^2 \rho=\frac{1}{r^2}-G(r)\left(\frac{G'(r)}{r G(r)}+\frac{1}{r^2}\right)
\end{equation}

From Fig. 11 we see that the energy density of dark matter at small distance is ill behaved in the case of NFW and  ''cusp'' phenomenon occurs. Interestingly, for the case of URC the dark matter density is finite at small distances. To conclude, the difference on the effect of dark matter on black hole shadow can be linked to the cusp phenomenon. For URC profile, the energy density of dark matter is finite near the black hole hence we observe smaller images.  This effect may be potentially explained by the fact that dark matter causes a pressure on the surrounding plasma near the black hole, as a result smaller shadow size is obtained.

\section{Radius distortion and energy emission}
In order to extract information the shadow size there are two commenly defined observable: the radius $R_s$ of the shadow and the distortion $\delta_s$. The shadow radius $R_s$ is defined in terms of the reference circle (see, Fig. 12) in connecting three charachteristic points on the boundary of the shadow: $(\alpha_\text{\footnotesize{t}},\beta_\text{\footnotesize{t}})$ corresponds to the top most point on the shadow,  $(\alpha_\text{\footnotesize{b}},\beta_\text{\footnotesize{b}})$ corrsponds to the bottom most point on the shadow and finally $(\alpha_\text{r},0)$ corrresponds to unstable circular orbit seen by an observer on reference frame. Mathematically the radius is approximated as follows \cite{pfdm}
\begin{equation}\label{observable}
R_s=\frac{(\alpha_\text{\footnotesize{t}}-\alpha_r)^2+\beta_\text{\footnotesize{t}}^2}{2|\alpha_{\text{\footnotesize{t}}}-\alpha_r|}.
\end{equation}

 \begin{figure}[h!]
  \includegraphics[width=0.44\textwidth]{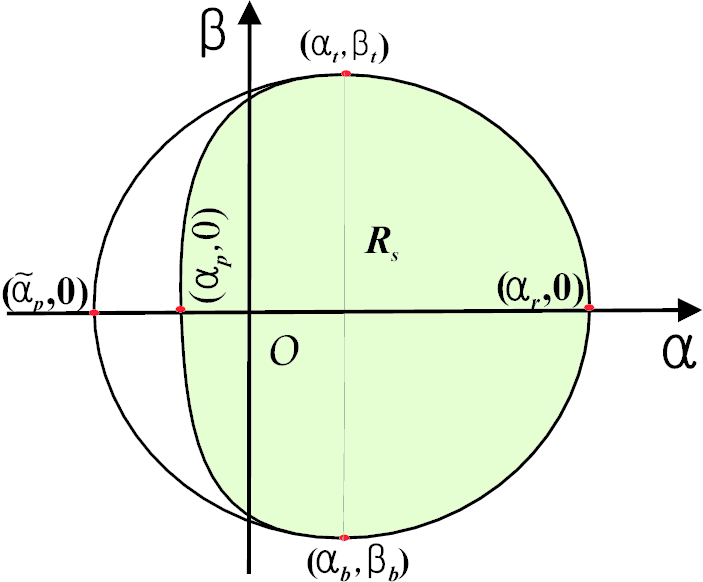}
  \caption{	Schematic representation of rotating black hole shadow and depicting the observable $R_s$ \cite{Amir2}.
  }
  \end{figure}
  
On the other hand, the second observable $\delta_s$ describes the rate of distortion. By construction $D_{CS}$ simply gives the difference between the contour of shadow and reference circle in terms of the points: $(\tilde{\alpha}_p,0)$ and $(\alpha_p,0)$, thus $D_{CS}=|\tilde{\alpha}_p-\alpha_p|$. Now the distortion yields
\begin{equation}
\delta_s=\frac{\tilde{\alpha}_p-\alpha_p}{R_s}.
\end{equation}

For our case, we consider the points $(\tilde{\alpha}_p,0)$ and $(\alpha_p,0)$ to be on the equatorial plane, opposite to the point $(\alpha_\text{r},0)$. 
   
    \begin{figure}[h!]
    \includegraphics[width=0.46\textwidth]{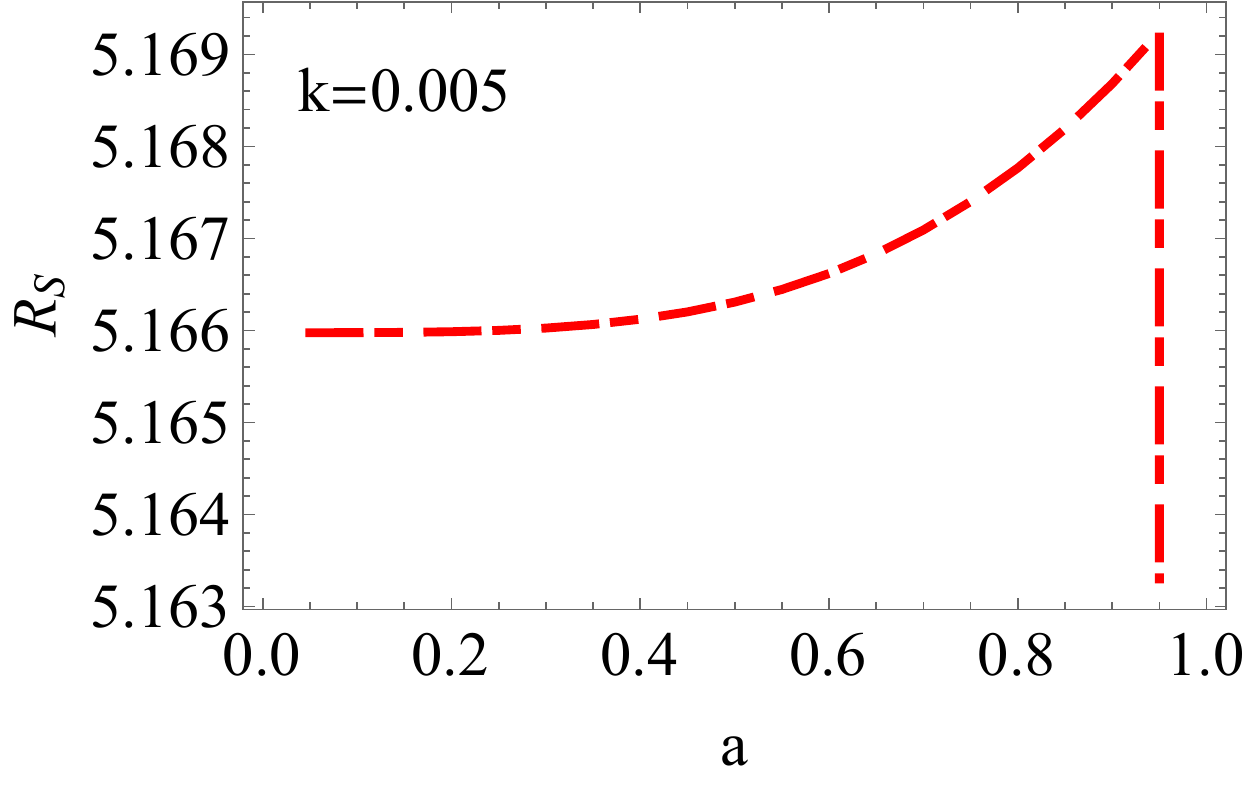}
        \includegraphics[width=0.45\textwidth]{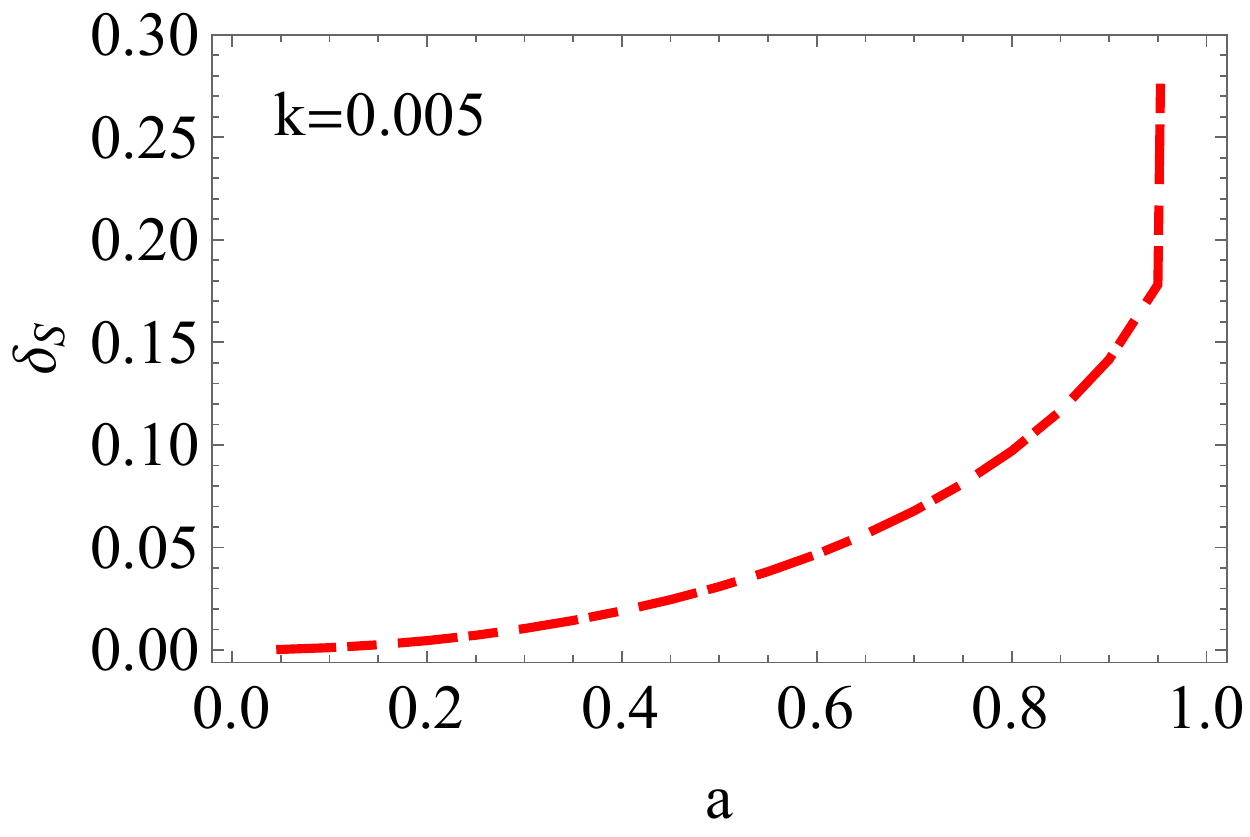}
  \caption{\label{figure1} Left panel: The quantities $R_s$  with respect to spin parameter $a$. Right panel: The quantities $\delta_s$  with respect to spin  parameter $a$.
  }
  \end{figure}

  \begin{figure}[h!]
    \includegraphics[width=0.42\textwidth]{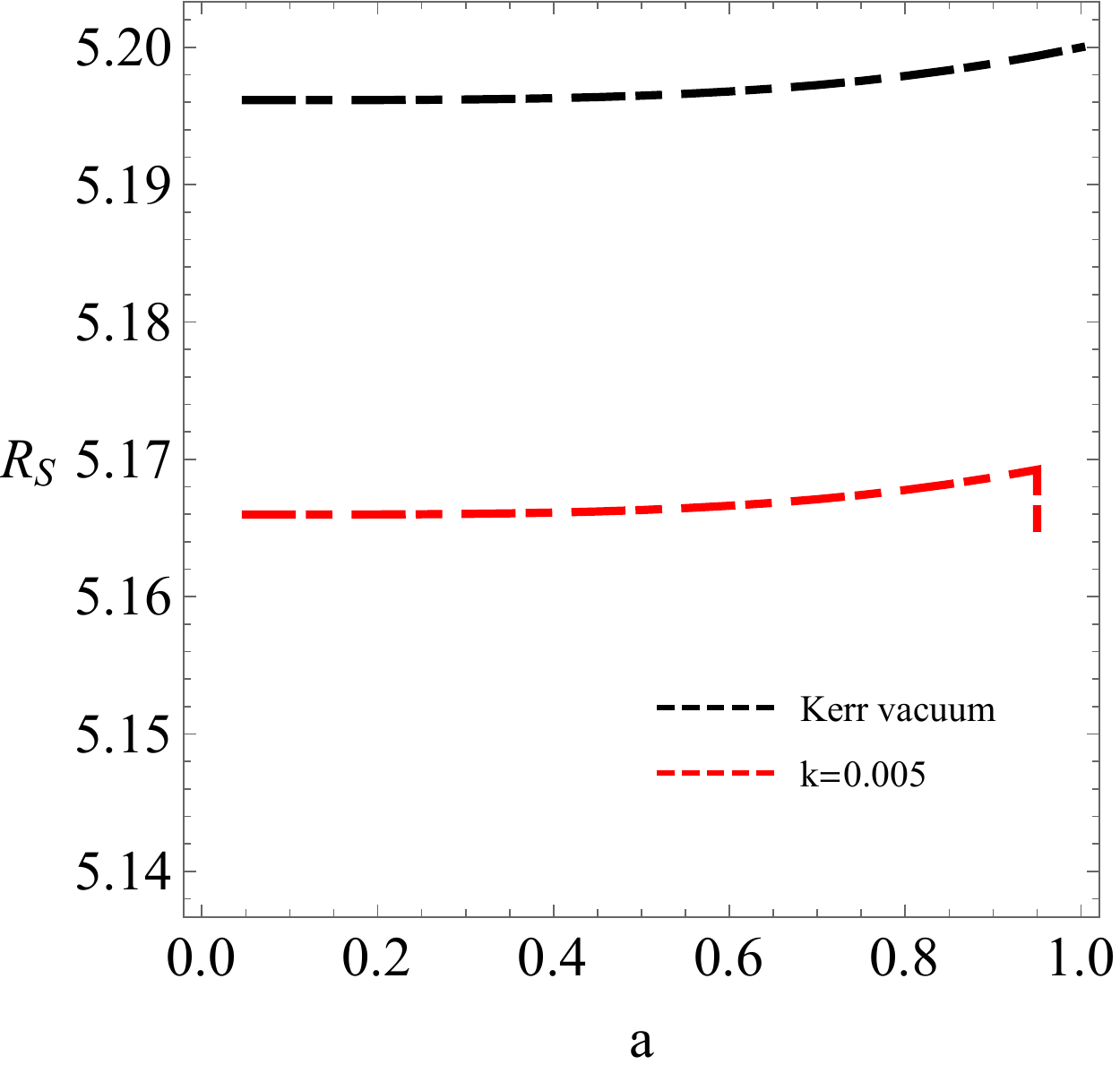}
        \includegraphics[width=0.42\textwidth]{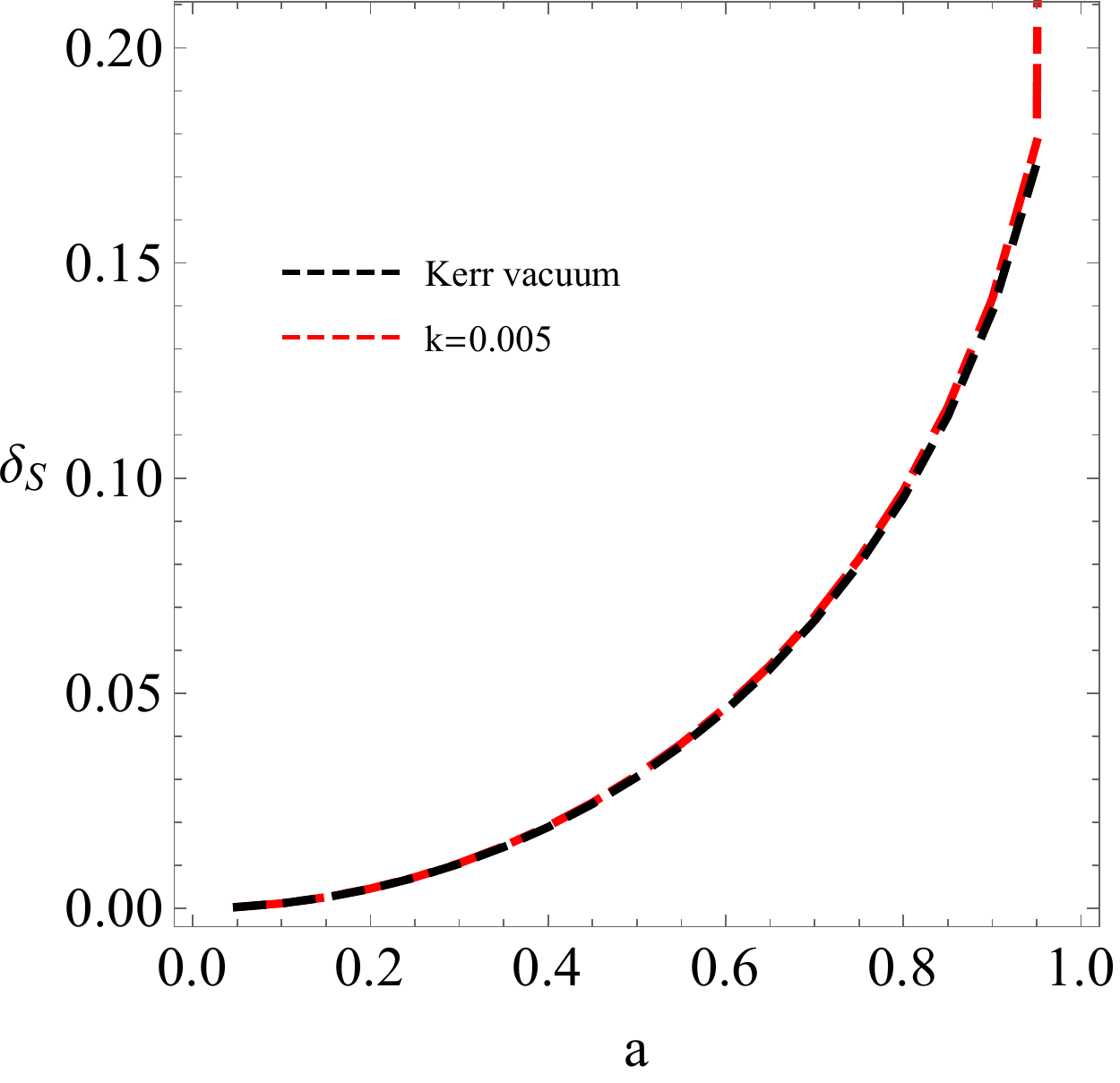}
  \caption{\label{figure1} Left panel: The quantities $R_s$  with respect to spin parameter $a$, The red curve corresponds to $k=0.005$, while the black curve to the case when dark matter effects are absent or almost negligible. Right panel: The quantities $\delta_s$  with respect to spin  parameter $a$. We see that the dark matter results in a decrease of the shadow radius.
  }
  \end{figure}
  
As we have considered our observer to be at infinity so in this case the area of the black hole shadow will be approximately equal to high energy absorption cross section. For a spherically symmetric black hole the absorption cross section oscillates around $\Pi_{ilm}$, a limiting constant value. For a black hole shadow with radius $R_s$, we adopt the value of $\Pi_{ilm}$  as calculated by 
\begin{equation}
  \Pi_{ilm}\approx ~ \pi R_s^2.
\end{equation}
The energy emission rate of the black hole is thus defined by \cite{pfdm}
\begin{equation}
\frac{d^2E(\sigma)}{d\sigma dt}=2 \pi ^2 \frac{\Pi_{ilm}}{e^{\sigma/T}-1}\sigma^3,
\end{equation}
where $\sigma$ is the frequency of the photon and $T$ represents the temperature of the black hole at outer horizon i.e. $r_+$, given by \cite{pfdm}
\begin{eqnarray}\notag
  T(r_+)&=&\lim_{r \to r_+}\frac{\partial_r\sqrt{g_{tt}}}{2 \pi \sqrt{g_{rr}}}\\
  &=&\left(2 a^2 \left(f(r_+)-1\right)+r_+(r_+^2+a^2)f^{'}(r_+)\right)\frac{r_+}{4 \pi \left(r_+^2+a^2\right)^2}
\end{eqnarray}

  \begin{figure}[h!]
    \includegraphics[width=0.46\textwidth]{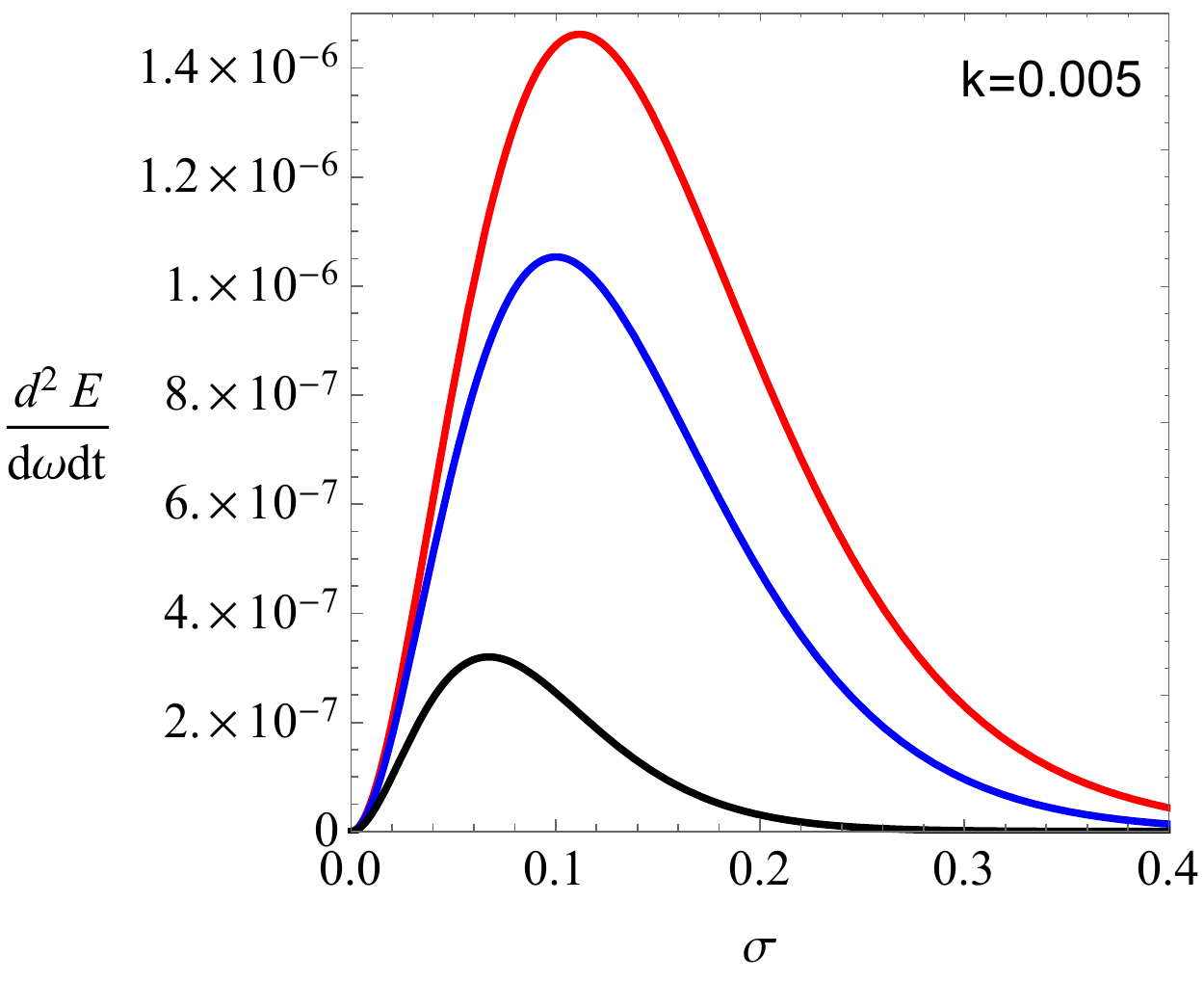}
        \includegraphics[width=0.46\textwidth]{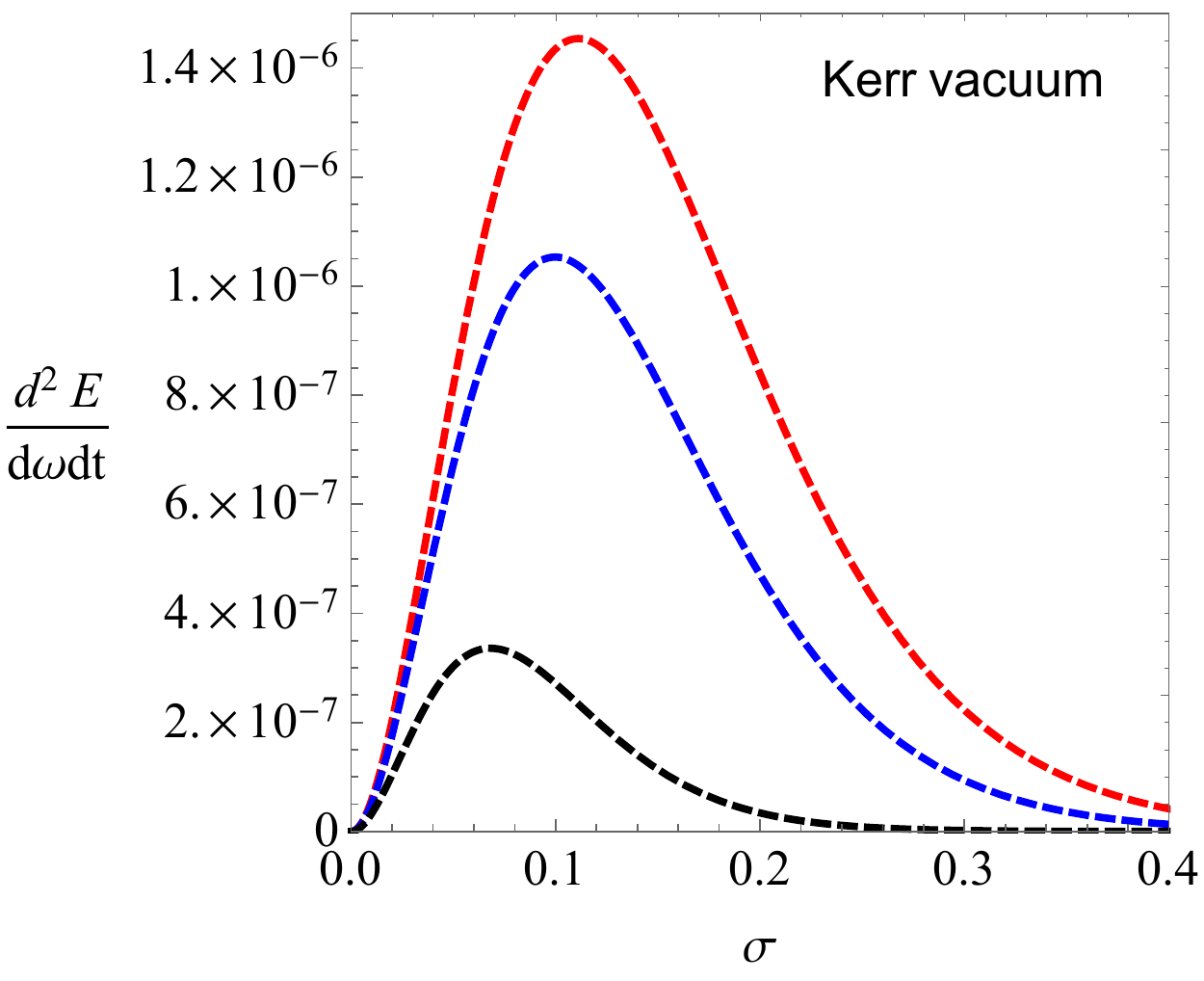}
  \caption{\label{figure1} Left panel: Emission rate for the case $k=0.005$. Right panel: Emission rate in the case when dark matter is absent. We use $a=0.2$ (red curve), $a=0.6$ (blue curve), and $a=0.9$ (black curve), respectively.
  }
  \end{figure}
  
The angular radius of the shadow can be estimated using the observable $R_s$ as $\theta_s = R_s M/D$, where $M$ is the black
hole mass and $D$ is the distance between the black hole and the observer. The angular radius can be further expressed
as $\theta_s = 9.87098 \times 10^{-6} R_s(M/$M\textsubscript{\(\odot\)})$(1kpc/ D)$ µas \cite{Hou:2018bar}. In the case of M87, for the  supermassive black hole M87 mass we have used $M = 6.5 \times  10^{9}$M\textsubscript{\(\odot\)} and $D =16.8$ Mpc \cite{m87} is the distance between the Earth and M87 center black hole.  The angular diameter of the shadow of the M87 supermassive black hole is estimated to be $ 42 \pm 3$ $\mu$as (see, \cite{m87}). In order to include dark matter effect, we have concidered the case of $k=0.005$  with $x=1$. We can estimate the observable quantities such as $R_s$ and $\delta_s$, as shown in Fig. 13. We we see from Fig. 14 that $R_s$ is smaller compared to the Kerr vacuum black hole when the dark matter effects are taken into account. Furthermore in Fig. 15 we plot the energy emission rate. Finally we compare our results with the empirical value of the angular diameter reported for the M87 estimated as $ 42$ $\mu$as (see, \cite{m87}), and found that our results are consistent with \cite{m87}. 

  \begin{figure}[h!]
    \includegraphics[width=0.45\textwidth]{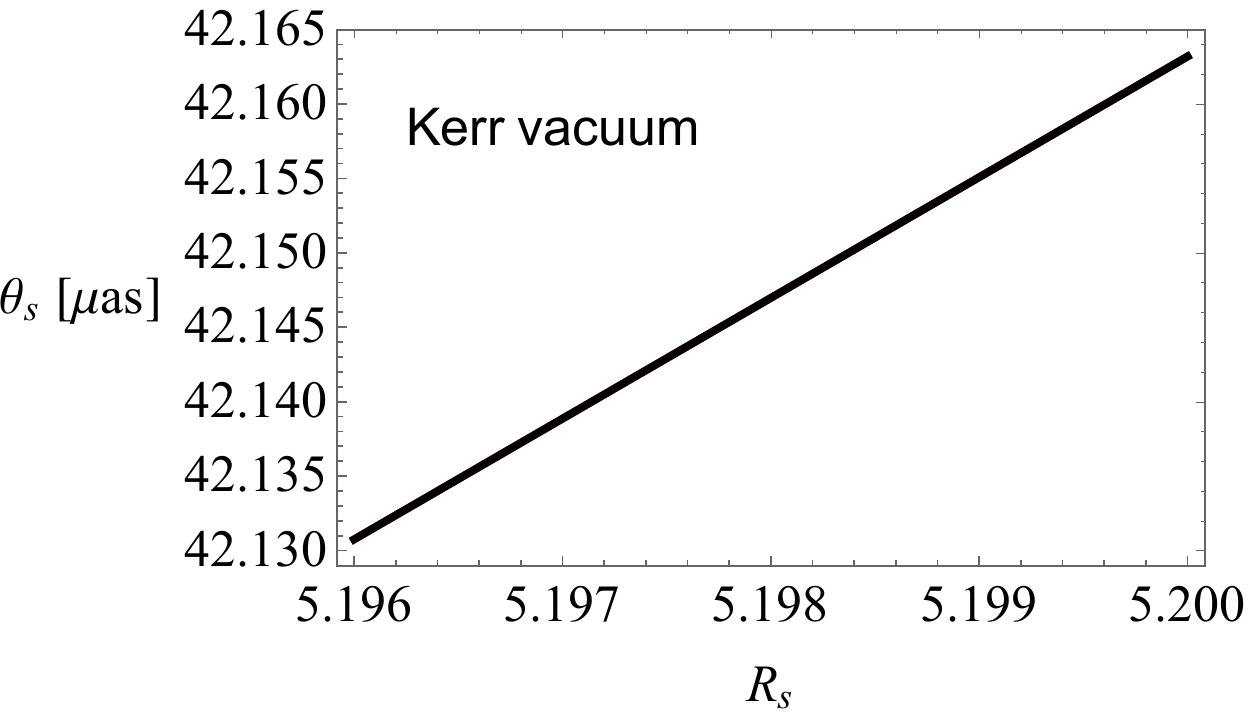}
        \includegraphics[width=0.45\textwidth]{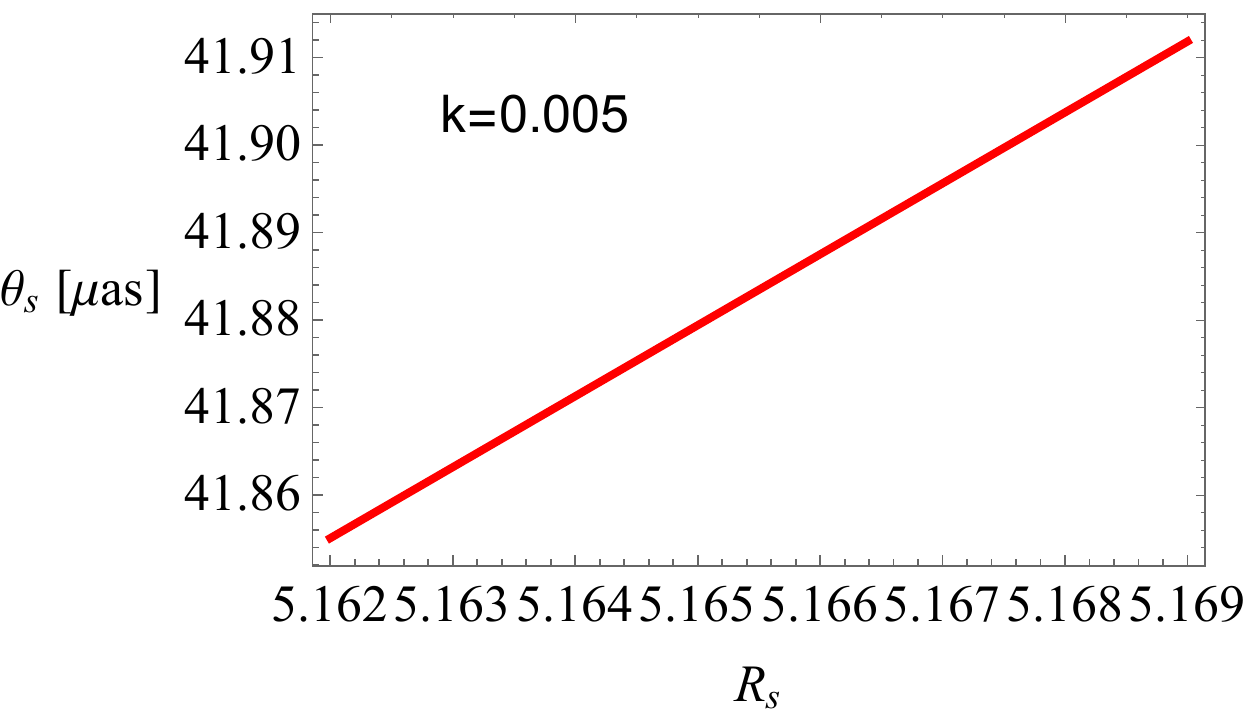}
  \caption{\label{figure1} Left panel: The angular diameter for M87 where the effect of dark matter are very small.  The consistency is achieved for the supermassive black hole M87 having mass $M = 6.9 \times  10^{9}$M\textsubscript{\(\odot\)}, hence the reported value of $ 42$ $\mu$as remains unchanged when we consider dark matter effects. Right panel: the angular diameter for the black hole with $k=0.005$ and $x=r/r_0=1$. Thus when the dark matter effects are stronger the angular diameters are shifted.  Note that we have adopted $D =16.8$ Mpc.
  }
  \end{figure}

\newpage
\section{Conclusion}
In this paper we have obtained a new  black hole solution surrounded by dark matter halo in the galactic center using the so-called Universal Rotation Curve (URC) dark matter profile.  Furthermore using the Newman-Jains method we have extended our solution to a  rotating black hole surrounded by dark matter halo.  We have explored the shadow images of the black hole M87 with effect of dark matter. In doing so, for the central density of the galaxy M87 we have adopted the value $\rho_0=6.9\,\times 10^{6}$ M\textsubscript{\(\odot\)}/kpc$^3$ and a core radius $r_0=91.2$ kpc. Our analyses show that the shadow size is the same compared to the Kerr vacuum solution. In the case of small spiral dominated by dark matter, such as UGC 7232, we find that the effect of dark matter on shadow images is almost same as in M87.  We have compared our results with \cite{m87} and found that the consistency is achieved if the black hole mass is $M = 6.9 \times  10^{9}$M\textsubscript{\(\odot\)}, hence the angular diameter reported for the M87 $ 42$ $\mu$as should remain almost unchanged in the presence of dark matter.  The small effect of dark matter on the shadow of suggests that most of the galactic dark matter resides in the halo and not in the core or center regions. Furthermore this show that it is very difficult to detect dark matter using the shadow images in terms of the present technology.  A similar results have been reportet recently in Ref. \cite{Hou:2018bar} where the effect of dark matter on the shadow of the Sgr A* was estimated to the order of $10^{-3} \mu$as. Although such tiny effects on the angular diameter are out of reach of the present technology,  it remains an open question if future astronomical observations can potentially detect such effects. \\

In order to have strong effect of dark matter we have argued that the core radius should be of the order of black hole masses implying a much higher value for the critical density, say compared to M87. In that case, the surrounding dark matter considerable affects the shadow shapes resulting with smaller angular diameter for the shadow.  This effect may be explained by the fact that dark matter causes a pressure on the surrounding plasma near the black hole, as a result smaller shadow size is obtained. We have compared two density profiles; the NFW and the URC dark matter density profile and obtained  opposite behavior on the black hole shadow. Namely, in the case of NFW the shadow radius increases, while for URC the shadow radius decreases. We have argued that this difference may be linked to the cusp phenomenon. In fact, the energy density of dark matter near the black hole is ill behaved while in the case of URC profile the energy density is finite. In that sense, the URC profile produces better results in our view. We plan in the near future to study the effect of stellar distribution on the black hole shadow.

\section*{Appendix A: Einstein field equations}

Here we shall show that the obtained spherical symmetric solution (37) is an exact solution of the Einstein field equations. It has been shown previously for the generic spherical symmetric solution generated from Newman-Janis algorithm does satisfy the Einstein field equations (see for details \cite{Azreg-Ainou:2014pra}). Toward this purpose can use the Einstein field equations $G_{\mu\nu}=8\pi T_{\mu\nu}$ along with the energy-momentum tensor represented by a properly chosen tetrad of the vector given by $T^{\mu \nu}=e^{\mu}_{a}e^{\nu}_{b}T^{ab}$, where $T^{ab}=(\rho,p_r,p_{\theta},p_{\varphi})$. In terms of the orthogonal basis, the energy momentum tensor components are given as follows
\begin{eqnarray}\nonumber
\rho &=&\frac{1}{8\pi}\emph{e}^\mu_t\,\emph{e}^\nu_t \,\emph{G}_{\mu\nu},\quad
p_r =\frac{1}{8\pi}\emph{e}^\mu_r\,\emph{e}^\nu_r \,\emph{G}_{\mu\nu},\\\label{m1}
p_\theta &=&\frac{1}{8\pi}\emph{e}^\mu_\theta\,\emph{e}^\nu_\theta \,\emph{G}_{\mu\nu},\quad
p_\varphi =\frac{1}{8\pi}\emph{e}^\mu_\varphi\, \emph{e}^\nu_\varphi\, \emph{G}_{\mu\nu}.
\end{eqnarray}

The idea is to find an orthogonal bases such that the Einstein field equations are satisfied
\begin{eqnarray}\notag
0&=&G_{\mu \nu}- 8\pi T_{\mu \nu}\\\notag
&=&G_{\mu \nu}-8\pi \left(e_{\mu}^{a}\,e_{\nu}^{b}\right)\frac{1}{8 \pi}\left(e^{\gamma}_{a}\,e^{\eta}_{b}\right)G_{\gamma \eta}\\\notag
&=& G_{\mu \nu}-(e_{\mu}^{a}\,e^{\gamma}_{a})\,(e_{\nu}^{b}\,e^{\eta}_{b})\,G_{\gamma \eta}\\\notag
&=& G_{\mu \nu}-\delta^{\gamma}_{\mu}\,\delta^{\eta}_{\nu}\,G_{\gamma \eta}\\
&=&0.
\end{eqnarray}

If such a bases exists, then the problem is solved. One such orthogonal bases is the following choice
\begin{eqnarray}\label{basis}
{\emph{e}}^\mu_t&=&\frac{1}{\sqrt{\Sigma \Delta}}\left(r^2+a^2,0,0,a\right),\quad
\emph{e}^\mu_r=\frac{\sqrt{\Delta}}{\sqrt{\Sigma}}\left(0,1,0,0\right),\\\nonumber
\emph{e}^\mu_\theta&=&\frac{1}{\sqrt{\Sigma}}\left(0,0,1,0\right),\quad
\emph{e}^\mu_\varphi=\frac{1}{\sqrt{\Sigma} \sin\theta}\left(a \sin^2\theta,0,0,1\right).
\end{eqnarray}

For the Einstein tensor components we find
\begin{eqnarray}\nonumber
\emph{G}_{tt}&=&\frac{2 \Upsilon' (a^4 \cos^4\theta-a^4\cos^2\theta+a^2 r^2+r^4-2 \Upsilon r^3)-a^2 r \sin^2\theta \Upsilon''}{\Sigma^3},\\\nonumber
\emph{G}_{rr}&=&-\frac{2 \Upsilon' r^2}{\Delta \Sigma},\\\label{einstein}
\emph{G}_{\theta\theta}&=& -\frac{\Upsilon'' a^2 r^2 \cos^2\theta+2\Upsilon' a^2 \cos^2\theta+\Upsilon'' r^3}{\Sigma},\\\nonumber
\emph{G}_{t\varphi}&=&\frac{a \sin^2\theta \left[r(a^2+r^2)\Sigma \Upsilon''+2 \Upsilon' \left((a^2+r^2)a^2\cos^2\theta-a^2 r^2-r^3(r-2 \Upsilon)\right)  \right]}{\Sigma^3},\\\nonumber
\emph{G}_{\varphi\varphi}&=&- \frac{\sin^2\theta\left[r(a^2+r^2)^2\Sigma\Upsilon''+2 a^2 \Upsilon' \left(\cos^2\theta (a^4+3a^2r^2+2 r^4-2 \Upsilon r^3)-a^2r^2-r^4+2\Upsilon r^3 \right)   \right]}{\Sigma^3}
\end{eqnarray}

Finally for the energy momentum tensor components we find
\begin{eqnarray}
\rho &=&\frac{2\Upsilon'(r)r^2}{8 \pi \Sigma^2}=-p_r,\,\,\,p_{\theta}=p_{\phi}=p_r-\frac{\Upsilon''(r)r+2\Upsilon'(r)}{8 \pi \Sigma}.
\end{eqnarray}

This shows that indeed the metric (37) is a solution of Einstein field equations. It's worth noting that our solution relies on the empirical dark matter density profile obtained from observations along with the rotating metric obtained by the Newman-Janis algorithm, a quite similar approach developed recently by Z.~Xu et al. \cite{Xu:2018wow}. One can go one step further to ask whether we can check the validity of the resulting energy momentum tensor components of the dark matter in the framework of some underlying field equations of the theory, say in terms of the explicit Lagrangian for the dark matter. Unfortunately, we don't know yet the full mechanism of the dark matter and this remains an open problem.  Starting from the action, equation of motion, and Lagrangian of the dark matter, in principle, one should be able to recover the above energy momentum tensor components.  Of course, this is outside the scope of the present article, but one possible way to tackle this problem is the idea of dark matter as a Bose-Einstein condensates (BEC). In particular the URC density profile obtained through the kinematics of spirals (i.e. observations) could be explained by the non-minimal coupling of a condensed phase inside a dark matter halo as an effective density profile (see for example, D.~Bettoni, et al. \cite{stefano}). According to this view, in large scales the dark matter is described by a minimally coupled weakly self-interacting scalar field, while inside the halos it develops a non-minimal coupling mechanism.

\end{document}